\documentclass
[
    a4paper,
    DIV=11,
    abstracton
]
{scrartcl}

\usepackage
{
    graphicx,
    amssymb,
    amsmath,
    amsthm,
    algorithm,
    algpseudocode,
    stmaryrd, 
    multirow,
    pbox,
    dsfont, 
    authblk,
    xcolor,
    enumitem
}

\usepackage[pdffitwindow=false,
            plainpages=false,
            pdfpagelabels=true,
            pdfpagemode=UseOutlines,
            pdfpagelayout=SinglePage,
            bookmarks=false,
            colorlinks=true,
            hyperfootnotes=false,
            linkcolor=blue,
            urlcolor=blue!30!black,
            citecolor=green!50!black]{hyperref}

\usepackage[bf]{caption}
\DeclareMathAlphabet{\mathpzc}{OT1}{pzc}{m}{it}

\newcommand{\subfiguretitle}[1]{{\scriptsize{#1}} \\[1ex]}
\newcommand{\R}{\mathbb{R}}
\newcommand{\mc}[1]{\mathpzc{#1}}

\renewcommand{\iff}{\; \Leftrightarrow \;}

\providecommand{\abs}[1]{\left\lvert #1 \right\rvert}
\providecommand{\norm}[1]{\left\lVert #1 \right\rVert}

\newcommand\xqed[1]{\leavevmode\unskip\penalty9999 \hbox{}\nobreak\hfill \quad\hbox{#1}}
\newcommand{\exampleSymbol}{\xqed{$\triangle$}}

\DeclareMathOperator{\diag}{diag}
\DeclareMathOperator{\tr}{tr}
\DeclareMathOperator{\Aut}{Aut}

\newtheorem{theorem}{Theorem}[section]

\newtheorem{lemma}[theorem]{Lemma}

\newtheorem{definition}[theorem]{Definition}
\theoremstyle{definition}
\newtheorem{example}[theorem]{Example}

\title{A Spectral Assignment Approach for the \\ Graph Isomorphism Problem}
\author[1]{Stefan Klus}
\author[2]{Tuhin Sahai}
\affil[1]{\normalsize Department of Mathematics and Computer Science, Freie Universit\"at Berlin, Germany}
\affil[2]{United Technologies Research Center, Berkeley, CA, USA}
\date{}

\begin{document}
\maketitle

\begin{abstract}
In this paper, we propose algorithms for the graph isomorphism (GI) problem that are based on the eigendecompositions of the adjacency matrices. The eigenvalues of isomorphic graphs are identical. However, two graphs $ \mc{G}_A $ and $ \mc{G}_B $ can be isospectral but non-isomorphic. We first construct a graph isomorphism testing algorithm for friendly graphs and then extend it to unambiguous graphs. We show that isomorphisms can be detected by solving a linear assignment problem. If the graphs possess repeated eigenvalues, which typically correspond to graph symmetries, finding isomorphisms is much harder. By repeatedly perturbing the adjacency matrices and by using properties of eigenpolytopes, it is possible to break symmetries of the graphs and iteratively assign vertices of $ \mc{G}_A $ to vertices of $ \mc{G}_B $, provided that an admissible assignment exists. This heuristic approach can be used to construct a permutation which transforms $ \mc{G}_A $ into $ \mc{G}_B $ if the graphs are isomorphic. The methods will be illustrated with several guiding examples.
\end{abstract}

\pagestyle{myheadings}
\thispagestyle{plain}

\section{Introduction}
\label{sec:Introduction}

We consider the problem of determining whether two undirected weighted graphs are isomorphic using spectral information. Efficient algorithms for the solution of the graph isomorphism or graph matching problem are required in a wide variety of different areas such as pattern recognition, object detection, image indexing, face recognition, and fingerprint analysis. Furthermore, novel applications such as the analysis of neural networks and social networks require matching of graphs with millions or even billions of vertices \cite{ABK15}. There exists no polynomial-time algorithm to check whether two arbitrary graphs are isomorphic. Interestingly, the graph isomorphism problem is one of only a few problems for which the complexity class is unknown~\cite{Koe06}. Although several attempts have been made to develop polynomial-time algorithms for the graph isomorphism problem, its complexity is currently unknown. Recently, Babai showed that GI can be solved in quasi-polynomial time \cite{Bab15}. GI belongs to the larger family of isomorphism problems on algebraic structures such as groups or rings that seem to lie between \textsl{P} and \textsl{NP}-complete~\cite{VT05}. Another open question is whether GI can be solved efficiently using quantum computers. The graph isomorphism problem can be regarded as a non-Abelian hidden subgroup problem (HSP) where the hidden subgroup is the automorphism group of the graph. An efficient solution of the HSP, which is the basis of many quantum algorithms, is only known for certain Abelian groups, whereas the general non-Abelian case remains open~\cite{HRT03}. For several special classes of graphs, however, the graph isomorphism problem is known to be solvable in polynomial time. These classes include, for instance, planar graphs~\cite{HT72} and graphs with bounded degree~\cite{Luk82} or bounded eigenvalue multiplicity~\cite{BGM82}. For an overview of isomorphism testing methods for these restricted graph classes, we refer to~\cite{Koe06}.

Since the graph isomorphism problem is challenging from a computational point of view, one is often forced to use different relaxations~\cite{FS15}. In \cite{ABK15}, the proposed graph isomorphism testing approach for friendly graphs is based on convex relaxations where the set of permutation matrices is replaced by the set of doubly stochastic matrices. First, a quadratic problem is solved to find a doubly stochastic matrix that minimizes the cost function; the solution is then, in a second step, projected onto the set of permutation matrices by solving a linear assignment problem. These results are extended to a larger class of graphs in~\cite{FS15}. In this paper, instead of a convex relaxation of the graph isomorphism problem, we consider a different relaxation where the set of permutation matrices is replaced by the set of orthogonal matrices. We then construct a linear assignment problem based on the eigenvectors of the graphs. We show that for a certain class of graphs, the solution of the linear assignment problem is also the unique solution of the graph isomorphism problem. For highly symmetric graphs, we propose an iterative algorithm which is based on spectral information and uses local perturbations of the adjacency matrices to break symmetries and to identify possible assignments. Our algorithm extends the applicability of existing spectral methods for the graph isomorphism problem to strongly regular graphs with repeated eigenvalues.

The paper is organized as follows: Section~\ref{sec:The graph isomorphism problem} contains a brief description of different formulations of the graph isomorphism problem. An overview of spectral properties of graphs is presented in Section~\ref{sec:Graphs with distinct eigenvalues}. Furthermore, we show how these properties can be used to find isomorphisms between graphs with simple spectrum. In Section~\ref{sec:Graphs with repeated eigenvalues}, we propose a novel eigendecomposition-based algorithm to determine whether two highly symmetric graphs are isomorphic. Numerical results for a number of different benchmark problems including strongly regular graphs and isospectral but non-isomorphic graphs are presented in Section~\ref{sec:Benchmark problems}. Section~\ref{sec:Conclusion} lists open questions and possible future work.

\section{The graph isomorphism problem}
\label{sec:The graph isomorphism problem}

Given two weighted undirected graphs $ \mc{G}_A = (\mc{V}, \mc{E}_A) $ and $ \mc{G}_B = (\mc{V}, \mc{E}_B) $ with adjacency matrices $ A $ and $ B $, where $ \mc{V} = \{ \mc{v}_1, \dots, \mc{v}_n \} $ is the set of vertices and $ \mc{E}_A $ and $ \mc{E}_B $ are the sets of edges, we want to determine whether these graphs are isomorphic.

\begin{definition}
Two graphs $ \mc{G}_A $ and $ \mc{G}_B $ are said to be \emph{isomorphic} -- denoted by $ \mc{G}_A \cong \mc{G}_B $~-- if one of the following equivalent conditions is satisfied:
\begin{enumerate}[label=\roman*)]
\item There exists a permutation $ \pi \in \mathcal{S}_n $ such that
\begin{equation*}
    (\mc{v}_i, \mc{v}_j) \in \mc{E}_A \iff (\mc{v}_{\pi(i)}, \mc{v}_{\pi(j)}) \in \mc{E}_B,
\end{equation*}
where $ \mathcal{S}_n $ denotes the symmetric group of degree $ n $.
\item There exists a permutation matrix $ P \in \mathcal{P}_n $ such that
\begin{equation*}
    B = P^T A P,
\end{equation*}
where $ \mathcal{P}_n $ is the set of all $ n \times n $ permutation matrices.
\end{enumerate}
\end{definition}

The relation between the permutation $ \pi $ and the permutation matrix $ P = (p_{ij}) $ is given by
\begin{equation*}
    p_{ij} =
    \begin{cases}
        1, & \text{if } \pi(i) = j, \\
        0, & \text{otherwise}.
    \end{cases}
\end{equation*}
The graph isomorphism problem can also be rewritten as a combinatorial optimization problem of the form
\begin{equation} \label{eq:GIP}
    \min_{P \in \mathcal{P}_n} \norm{ B - P^T A P }_F,
\end{equation}
where $ \norm{.}_F $ denotes the Frobenius norm. The graphs $ \mc{G}_A $ and $ \mc{G}_B $ are isomorphic if and only if the minimum of the above cost function is zero. Since
\begin{equation*}
    \norm{B - P^T A P}_F^2 = \norm{B}_F^2 - 2 \tr\left(B^T P^T A P \right) + \norm{A}_F^2,
\end{equation*}
cost function~\eqref{eq:GIP} is minimized if the term $ \tr\left( B^T P^T A P \right) $, which is the cost function of the NP-complete quadratic assignment problem (QAP), is maximized and vice versa.

\begin{definition}
An isomorphism from a graph $ \mc{G} $ to itself is called an \emph{automorphism}. 
\end{definition}

The set of all automorphisms of a graph forms a group under matrix multiplication, the so-called \emph{automorphism group}, typically denoted by $ \Aut(\mc{G}) $. If the graphs $ \mc{G}_A $ and $ \mc{G}_B $ are isomorphic, then the number of isomorphisms from $ \mc{G}_A $ to $ \mc{G}_B $ is identical to the number of automorphisms of $ \mc{G}_A $ or $ \mc{G}_B $, respectively~\cite{Spi09}.

\begin{definition}
A graph $ \mc{G} $ is called \emph{asymmetric} if the automorphism group is trivial.
\end{definition}

That is, $ \Aut(\mc{G}) = \{ I \} $ for asymmetric graphs. If the automorphism group is nontrivial, we call the graph \emph{symmetric}. The automorphism groups of strongly regular graphs, for example, can be highly nontrivial.

\begin{definition}
A graph is said to be \emph{regular} (or \emph{weakly regular}) if each vertex has the same number of neighbors and \emph{strongly regular} if additionally integers $ \alpha $ and $ \beta $ exist such that every pair of vertices $ \mc{v}_i $ and $ \mc{v}_j $ shares exactly $ \alpha $ common neighbors if the vertices $ \mc{v}_i $ and $ \mc{v}_j $ are adjacent and exactly $ \beta $ common neighbors otherwise \cite{Spi96}.
\end{definition}

The Frucht graph shown in Figure~\ref{fig:Simple graphs}a, for instance, is weakly regular. An example of a strongly regular graph is the Paley graph shown in Figure~\ref{fig:Paley}a. The solution of the minimization problem \eqref{eq:GIP} is not unique if the graphs possess nontrivial symmetries; for isomorphic graphs, the number of feasible solutions corresponds to the number of isomorphisms. Whether symmetries exist or not, however, is in general difficult to determine a priori. Although symmetries typically correspond to repeated eigenvalues, the correspondence is not exact~\cite{Lov07}. Examples of asymmetric graphs with repeated eigenvalues or symmetric graphs with simple spectrum (see also Figure~\ref{fig:Simple graphs}c) can be found in~\cite{FS15}, for example. In what follows, we will characterize graphs using spectral properties.

\section{Graphs with distinct eigenvalues}
\label{sec:Graphs with distinct eigenvalues}

An isomorphism testing algorithm for graphs with $ n $ distinct eigenvalues developed by Leighton and Miller is presented in~\cite{Spi09, LM79}. The method determines the isomorphism by breaking vertices that are not equivalent into different classes. Based on the entries of the eigenvectors, these classes are refined until either an isomorphism is found or the graphs are shown to be non-isomorphic. Another approach that utilizes spectral information is presented in~\cite{ABK15}, where so-called friendly graphs (defined below) are considered. The set of permutation matrices is replaced by the set of doubly stochastic matrices. After solving the resulting quadratic program, a linear assignment problem is solved to project the doubly stochastic matrix back onto the set of permutation matrices. Aflalo et al.\ \cite{ABK15} prove that for friendly isomorphic graphs, the relaxed problem is equivalent to the original graph isomorphism problem. We will use a different approach for graph isomorphism testing that relies on a relaxation to the manifold of orthogonal matrices and -- for friendly graphs -- requires only the solution of a single linear assignment problem. An extension of this method for graphs with repeated eigenvalues will be proposed in Section~\ref{sec:Graphs with repeated eigenvalues}.

\subsection{The two-sided orthogonal Procrustes problem}

Let $ \mathcal{O}_n = \left\{ P \in \R^{n \times n} \mid P^T P = I \right\} $ denote the set of all orthogonal matrices. Note that the set of permutation matrices $ \mathcal{P}_n $ is a subset of $ \mathcal{O}_n $. Provided that the matrices $ A $ and $ B $ are symmetric\footnote{The symmetry of the adjacency matrix should not be confused with the aforementioned graph symmetries.} -- we consider only undirected graphs --, the solution of the relaxed problem
\begin{equation} \label{eq:RGIP}
    \min_{P \in \mathcal{O}_n} \norm{ B - P^T A P }_F,
\end{equation}
which is called \emph{two-sided orthogonal Procrustes problem}~\cite{Sch68, GD04}, can be computed analytically, provided that both $ A $ and $ B $ have distinct eigenvalues. This result is captured in the following theorem:

\begin{theorem} \label{thm:Procrustes}
Given two symmetric matrices $ A $ and $ B $ with distinct eigenvalues, let $ A = V_A \Lambda_A V_A^T $ and $ B = V_B \Lambda_B V_B^T $ be the eigendecompositions, with $ \Lambda_A = \diag\big(\lambda_A^{(1)}, \dots, \lambda_A^{(n)}\big) $, $ \Lambda_B = \diag\big(\lambda_B^{(1)}, \dots, \lambda_B^{(n)}\big) $, and $ \lambda_A^{(1)} < \dots < \lambda_A^{(n)} $ as well as $ \lambda_B^{(1)} < \dots < \lambda_B^{(n)} $. Then the orthogonal matrix $ P^* $ which minimizes \eqref{eq:RGIP} is given by
\begin{equation*}
    P^* = V_A S V_B^T,
\end{equation*}
where $ S = \diag(\pm 1, \dots, \pm 1) $.
\end{theorem}

A proof of the above theorem can be found in \cite{Sch68}. The eigenvalues and corresponding eigenvectors have to be sorted both in either increasing or decreasing order. Note that there are $ 2^n $ different orthogonal matrices which minimize the cost function.

\begin{lemma}\label{lem:ProcrustesPerm}
If the graphs $ \mc{G}_A $ and $ \mc{G}_B $ are isomorphic, then one of the $ 2^n $ solutions is the permutation matrix $ P $ that minimizes \eqref{eq:GIP}.
\end{lemma}
\begin{proof}
Since all eigenvalues are distinct, the eigenvectors of $ B $ are -- up to the signs\footnote{The eigenvectors are, without loss of generality, assumed to be normalized.} -- permutations of the eigenvectors of $ A $, i.e., $ V_B = P V_A \hat{S} $, where $ P \in \mathcal{P}_n $ permutes the rows and $ \hat{S} = \diag(\pm 1, \dots, \pm 1) $ flips the signs of the eigenvectors. If we now choose $ S = \hat{S} $, then $ P^* = P V_A \hat{S}^2 V_A^T $. Using $ \hat{S}^2 = I $ and the orthogonality of $ V_A $, we obtain $ P^* = P \in \mathcal{P}_n $.
\end{proof}

Note that we are, however, not searching over the $ 2^n $ solutions. Let now $ c $ be the cost of the optimal solution of the relaxed problem, i.e.
\begin{equation*}
    c = \min_{P \in \mathcal{O}_n} \norm{ B - P^T A P }_F
      = \norm{ B - P^{*T} A P^* }_F
      = \norm{ \Lambda_B - \Lambda_A }_F.
\end{equation*}
As $ \mathcal{P}_n \subset \mathcal{O}_n $, the graphs cannot be isomorphic if $ c \ne 0 $. If, on the other hand, $ c = 0 $, this implies that the eigenvalues of $ A $ and $ B $ are identical and the graphs $ \mc{G}_A $ and $ \mc{G}_B $ are isospectral but not necessarily isomorphic. In addition to the eigenvalues, the eigenvectors of the graphs can be used for isomorphism testing as shown in the following example.

\begin{example}
An example of isospectral graphs, taken from~\cite{OB12}, is shown in Figure~\ref{fig:Isospectral}. Setting $ a = 1 $, $ b = 2 $, and $ c = 3 $, the eigenvectors of the graphs belonging to the largest eigenvalue $ \lambda_A^{(6)} = \lambda_B^{(6)} = 5.167 $ are
\begin{equation*}
    \begin{split}
        v_A^{(6)} &= [0.380, 0.092, 0.157, 0.655, 0.477, 0.407]^T, \\
        v_B^{(6)} &= [0.222, 0.068, 0.352, 0.575, 0.353, 0.606]^T.
    \end{split}
\end{equation*}
Since the entries of the eigenvectors are different, $ v_B^{(6)} $ cannot be written as a permutation of $ v_A^{(6)} $, implying that $ \mc{G}_A $ and $ \mc{G}_B $ are not isomorphic. \exampleSymbol

\begin{figure}[htb]
    \centering
    \begin{minipage}[t]{0.4\textwidth}
        \centering
        \subfiguretitle{a)}
        \includegraphics[width=0.8\textwidth]{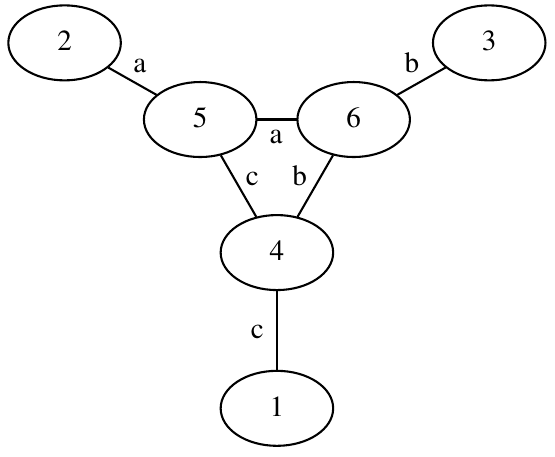}
    \end{minipage}
    \begin{minipage}[t]{0.4\textwidth}
        \centering
        \subfiguretitle{b)}
        \includegraphics[width=0.8\textwidth]{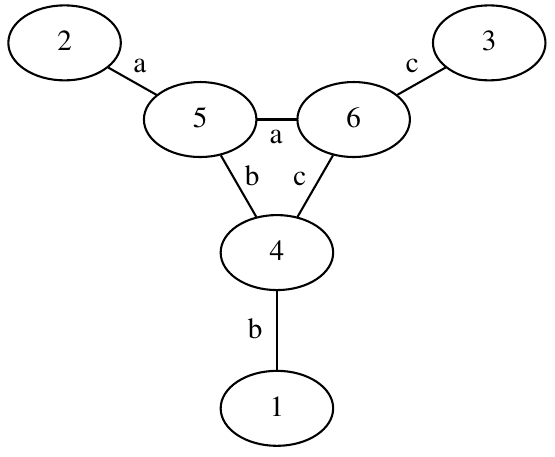}
    \end{minipage}
    \caption{Isospectral graphs.}
    \label{fig:Isospectral}
\end{figure}

\end{example}

Provided that the entries of the eigenvectors are distinct, this simple test can be used for graph isomorphism testing. Here, we compared the entries of the eigenvectors corresponding to the largest eigenvalue since the largest eigenvalue of the adjacency matrix of a connected graph always has multiplicity one~\cite{Lov07}. Furthermore, all entries of the corresponding eigenvector are strictly positive. Note that in general the normalized eigenvectors are only determined up to the sign. Thus, two comparisons might be required. If the absolute values of all entries of an eigenvector are different -- Spielman~\cite{Spi09} calls such an eigenvector \emph{helpful} --, this gives us a canonical labeling of the vertices and the graph isomorphism problem can be solved easily.

\subsection{Friendly and unambiguous graphs}

In \cite{ABK15}, friendly graphs are considered. In contrast to asymmetry, friendliness can be easily verified.

\begin{definition} \label{def:Friendly graph}
Let $ \mathds{1} \in \R^n $ denote the vector of ones. A graph $ \mc{G}_A $ with adjacency matrix $ A $ is called \emph{friendly} if $ A $ has distinct eigenvalues and all eigenvectors $ v_A^{(k)} $ satisfy $ \mathds{1}^T v_A^{(k)} \ne 0 $.
\end{definition}

As a result, it is possible to make the signs of corresponding eigenvectors consistent\footnote{E.g., by ensuring that for all eigenvectors $ \mathds{1}^T v_A^{(k)} > 0 $ and $ \mathds{1}^T v_B^{(k)} > 0 $.}. This corresponds to finding the sign matrix $ S $ in Lemma~\ref{lem:ProcrustesPerm}. Thus, the permutation matrix which solves the graph isomorphism problem can be computed directly. We will propose a different approach that relies on the solution of a linear assignment problem and will be generalized later on. For friendly graphs, we obtain:

\begin{lemma}
Every friendly graph $ \mc{G}_A $ is asymmetric~\cite{ABK15}.
\end{lemma}

\begin{proof}
Let $ A = V_A \Lambda_A V_A^T $ be the eigendecomposition. Assuming there exists $ P \in \mathcal{P}_n $ with $ A = P^T A P $, we obtain another eigendecomposition $ A = (P^T V_A) \Lambda_A (P^T V_A)^T $. Since the eigenvectors are determined up to the signs, $ V_A = P^T V_A S $, where $ S = \diag(\pm 1, \dots, \pm 1) $. Thus, $ \mathds{1}^T V_A = \mathds{1}^T P^T V_A S = \mathds{1}^T V_A S $. Each entry of the vector $ \mathds{1}^T V_A $ must be nonzero since the graph is friendly. Thus, the equation can only be satisfied if $ S = I $. However, $ S = I $ implies $ V_A = P^T V_A $ and hence $ P = I $. As a result, the automorphism group only contains the identity matrix and the graph is asymmetric.
\end{proof}

\begin{figure}[htb]
    \centering
    \includegraphics[width=0.55\textwidth]{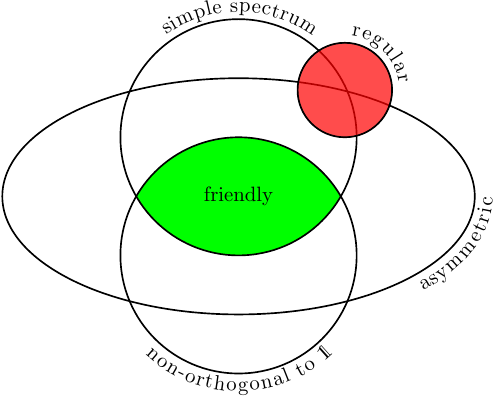}
    \caption{Different classes of graphs based on~\cite{FS15}. Friendly graphs are given by the intersection of graphs with simple spectrum and graphs whose eigenvectors are non-orthogonal to $ \mathds{1}$.}
    \label{fig:Venn}
\end{figure}

The converse is not true, the Frucht graph shown in Figure~\ref{fig:Simple graphs}a, for instance, is asymmetric but not friendly. Furthermore, there are asymmetric graphs with repeated eigenvalues. A Venn diagram of different graph classes is shown in Figure~\ref{fig:Venn} (reproduced from \cite{FS15}, examples of graphs in each of these categories can also be found there). Even if eigenvectors are not friendly, it is often possible to make the signs of two corresponding eigenvectors consistent.

\begin{definition}
We call an eigenvector $ v $ \emph{ambiguous} if $ v $ and $ -v $ have exactly the same entries. That is, there exists $ P \in \mathcal{P}_n $ with $ v = -P v $. A graph without ambiguous eigenvectors is called \emph{unambiguous}.
\end{definition}

This property can be easily verified by sorting the entries. Note that ambiguity implies unfriendliness since $ v = -P v $ results in $ \mathds{1}^T v = -\mathds{1}^T v $ and thus $ \mathds{1}^T v = 0 $. The class of graphs whose eigenvectors are not ambiguous is thus larger than the class of friendly graphs.

\subsection{A spectral assignment approach for friendly and unambiguous graphs}

We now want to construct a linear assignment problem for friendly graphs that solves the graph isomorphism problem. In our approach, the cost of assigning vertices of $ \mc{G}_A $ to vertices of $ \mc{G}_B $ will be based on the eigenvectors.

\begin{definition} \label{def:Cost matrix}
Let $ V, W \in \R^{n \times m} $ be two matrices, then the cost of assigning $ V $ to $ W $ is defined to be $ C(V, W) = (c_{ij}) \in \R^{n \times n} $, with 
\begin{equation*}
    c_{ij} = \sum_{k=1}^m \big\lvert v_i^{(k)} - w_j^{(k)} \big\rvert,
\end{equation*}
where $ v^{(k)} $ and $ w^{(k)} $ are the column vectors of $ V $ and $ W $, respectively.
\end{definition}

\begin{lemma} \label{lem:PermVecMat}
Let $ V $, $ W $, and $ C = C(V, W) $ be as in the above definition. Define
\begin{equation} \label{eq:LAP}
    c = \min_{P \in \mathcal{P}_n} \tr \left( P^T C \right)
\end{equation}
to be the minimal cost of the linear assignment problem. Then $ V = P W $ for $ P \in \mathcal{P}_n $ if and only if $ c = 0 $.
\end{lemma}

\begin{proof}
We will first show that this result holds for vectors $ v $ and $ w $. Assume that $ v = P w $ for $ P \in \mathcal{P}_n $. It follows that $ v_i = w_{\pi(i)} $ and thus $ c_{i, \pi(i)} = \abs{ v_i - w_{\pi(i)} } = 0 $. Furthermore,
\begin{equation*}
    \tr \left( P^T C \right) = \sum_{i,j=1}^n p_{ij} c_{ij} = \sum_{i=1}^n c_{i, \pi(i)} = 0.
\end{equation*}
For the other direction, assume that $ c = 0 $ and that the corresponding permutation matrix is $ \hat{P} $. Then $ c_{i, \hat{\pi}(i)} = \abs{ v_i - w_{\hat{\pi}(i)} } = 0 $ and consequently $ v = \hat{P} w $.
For matrices, the proof is almost identical. Assume that $ V = P W $. Then $ v^{(k)} = P w^{(k)} $ for all column vectors. Thus, with the first part, we obtain $ c_{i, \pi(i)} = 0 $ and $ \tr \left( P^T C \right) = 0 $. The other direction follows in the same way.
\end{proof}

The linear assignment problem can be solved in $ \mathit{O}(n^3) $ using the Hungarian method~\cite{Kuh55, BC99}. For two friendly graphs $ \mc{G}_A $ and $ \mc{G}_B $ with adjacency matrices $ A $ and $ B $ and eigendecompositions as described in Theorem~\ref{thm:Procrustes}, we define $ C(\mc{G}_A, \mc{G}_B) = C(V_A, V_B) $. Note that in the following theorem we assume that the signs of the corresponding eigenvectors $ v_A^{(k)} $ and $ v_B^{(k)} $ are consistent. This is possible as the sum of the entries of each eigenvector is nonzero.

\begin{theorem} \label{thm:GI friendly graphs}
Let $ \mc{G}_A $ and $ \mc{G}_B $ be friendly graphs. Define $ C = C(\mc{G}_A, \mc{G}_B) $ and $ c $ as in \eqref{eq:LAP}. Then $ \mc{G}_A \cong \mc{G}_B $ if and only if $ \Lambda_A = \Lambda_B $ and $ c = 0 $. The solution $ P $ of the linear assignment problem is then a solution of GI.
\end{theorem}

\begin{proof}
Assume that $ \mc{G}_A \cong \mc{G}_B $ and thus $ A = V_A \Lambda_A V_A^T = P B P^T = (P V_B) \Lambda_B (P V_B)^T $. Using Lemma~\ref{lem:PermVecMat}, we obtain $ \tr \left( P^T C \right) = 0 $.
If, on the other hand, $ c = 0 $, Lemma~\ref{lem:PermVecMat} implies that $ V_A = P V_B $ and thus
\begin{equation*}
    \norm{B - P^T A P}_F = \norm{V_B \Lambda_B V_B^T - P^T V_A \Lambda_A V_A^T P}_F = \norm{ \Lambda_B - \Lambda_A}_F = 0. \qedhere
\end{equation*}
\end{proof}

\begin{algorithm}[htb]
    \caption{Graph isomorphism testing for friendly graphs.}
    \label{alg:GI friendly graphs}
    \vspace*{-0.5\baselineskip}
    \begin{enumerate} \setlength{\itemsep}{0mm}
    \item Compute eigendecompositions of $ A $ and $ B $ as described in Theorem~\ref{thm:Procrustes}.
    \item Make signs of eigenvectors consistent so that $ \mathds{1}^T v_A^{(k)} > 0 $ and $ \mathds{1}^T v_B^{(k)} > 0$. (If the eigenvectors are ambiguous, we will use the absolute values of the entries.)
    \item Compute $ C = C(\mc{G}_A, \mc{G}_B) $ and $ c = \min\limits_{P \in \mathcal{P}_n} \tr \left( P^T C \right) $.
    \item If $ \Lambda_A = \Lambda_B $ and $ c = 0 $, then $ \mc{G}_A \cong \mc{G}_B $.
    \end{enumerate}
    \vspace*{-0.5\baselineskip}
\end{algorithm}

For friendly graphs, we propose the graph isomorphism testing approach described in Algorithm~\ref{alg:GI friendly graphs}. If an eigenvector is ambiguous, we will use the absolute values of the entries for the computation of the cost matrix.

\begin{example} \label{ex:Ambiguity}
Let us illustrate the definition of ambiguity\footnote{In what follows, we will sometimes use the shorter cycle notation for permutations. That is, a permutation is represented as a product of cycles, where cycles of length one are omitted. E.g., $ \pi = (1 \; 2 \; 3) $ means that $ 1 $ is assigned to $ 2 $, $ 2 $ to $ 3 $, and $ 3 $ to $ 1 $, while $ 4 $ remains unchanged.}:
\begin{enumerate}[label=\roman*)]
\item The vectors $ v = [1, \, 2, \, 0, \, -3]^T $ and $ w = [0, \, -1, \, -2, \, 3]^T $ are unfriendly but not ambiguous and $ v $ can only be assigned to $ -w $ using $ \pi = (1 \; 2 \; 3) $.
\item The vectors $ v = [1, \, 2, \, -1, \, -2]^T $ and $ w =[-2, \, -1, \, 1, \, 2]^T $, on the other hand, are ambiguous since $ v $ can be assigned to $ w $ using $ \pi = (1 \; 3 \; 2 \; 4) $ or to $ -w $ using $ \pi = (1 \; 2) $. Taking absolute values leads to two spurious solutions, given by $ \pi = (1 \; 3 \; 2) $ and $ \pi = (1 \; 2 \; 4) $. \exampleSymbol
\end{enumerate}
\end{example}

If the eigenvectors are not ambiguous, we can make the signs consistent (e.g., by sorting the entries) and apply Algorithm~\ref{alg:GI friendly graphs} in the same way by replacing only step 2. Note that we only assumed that the signs of the eigenvectors are consistent in Theorem~\ref{thm:GI friendly graphs}, the friendliness property was not used explicitly.

\begin{example} \label{ex:GI distint eigenvalues}
Let us consider different graph types to illustrate the idea behind the assignment approach:

\begin{enumerate}[leftmargin=0em,itemindent=1.7em,labelsep=0.3em,label=\roman*)] \setlength{\itemsep}{0mm}
\item Given the Frucht graph shown in Figure~\ref{fig:Simple graphs}a and the permutation of the graph shown in Figure~\ref{fig:Simple graphs}b, the resulting cost matrix $ C $ is displayed in Figure~\ref{fig:Simple graphs}g. The solution of the linear assignment problem is
\begin{equation*}
    \pi =
    \begin{bmatrix}
        1 & 2 & 3 & 4 & 5 & 6 & 7 & 8 & 9 & 10 & 11 & 12 \\
        4 & 5 & 6 & 1 & 2 & 3 & 9 & 8 & 7 & 12 & 11 & 10
    \end{bmatrix}
\end{equation*}
and the cost of the assignment is zero. The graph has simple spectrum, is asymmetric, regular, and thus not friendly\footnote{This is due to the fact that $ \mathds{1} $ is always an eigenvector of regular graphs, all the other eigenvectors must be perpendicular and are hence not friendly, cf.~\cite{FS15}.}. However, only one eigenvector is ambiguous. Even without taking into account this eigenvector, the algorithm successfully computes the correct permutation matrix.

\item An example of a graph with nontrivial automorphism group but simple spectrum, taken from \cite{FS15}, and a random permutation are shown in Figure~\ref{fig:Simple graphs}c--d, the corresponding cost matrix $ C $ is depicted in Figure~\ref{fig:Simple graphs}h. Here, two eigenvectors are ambiguous and the solution of the LAP is not unique since $ \mc{v}_1 $ could be assigned to $ \mc{v}_2 $ or $ \mc{v}_4 $ and $ \mc{v}_2 $ to $ \mc{v}_1 $ or $ \mc{v}_3 $. These assignments, however, are not independent, as soon as one of the first four vertices is assigned, the others follow automatically. Feasible solutions are
\begin{equation*}
    \pi_1 =
    \begin{bmatrix}
        1 & 2 & 3 & 4 & 5 & 6 & 7 & 8 \\
        4 & 3 & 2 & 1 & 6 & 5 & 7 & 8
    \end{bmatrix}, \quad
    \pi_2 =
    \begin{bmatrix}
        1 & 2 & 3 & 4 & 5 & 6 & 7 & 8 \\
        2 & 1 & 4 & 3 & 6 & 5 & 7 & 8
    \end{bmatrix}.
\end{equation*}
If we do not use the absolute values of the ambiguous eigenvectors for the computation of the cost matrix, then there are four possible combinations: We can assign $ v_A^{(3)} $ to $ \pm v_B^{(3)} $ and $ v_A^{(6)} $ to $ \pm v_B^{(6)} $. One combination will result in $ \pi_1 $, one in $ \pi_2 $, the remaining two lead to nonzero assignment costs. Thus, taking absolute values prevents conflicting information about possible assignments, but also results in spurious solutions (cf.~Example~\ref{ex:Ambiguity}).

\item The Facebook social circles graph~\cite{ML12}, available through the SNAP website~\cite{LK14}, consists of 4039 vertices and 88234 edges. The adjacency matrices of the original and permuted graph are shown in Figure~\ref{fig:Simple graphs}e--f. The cluster structure of the graph is clearly visible in Figure~\ref{fig:Simple graphs}e. Numerically, the graph has a couple of repeated eigenvalues (around $ \lambda = -1 $, $ \lambda = 0 $, and $ \lambda = 1 $) and we neglect the corresponding eigenvectors. Nevertheless, Algorithm \ref{alg:GI friendly graphs} returns a valid assignment that solves the graph isomorphism problem. \exampleSymbol
\end{enumerate}

\begin{figure}[htbp]
    \centering
    \begin{minipage}[t]{0.45\textwidth}
        \centering
        \subfiguretitle{a)}
        \includegraphics[width=0.85\textwidth]{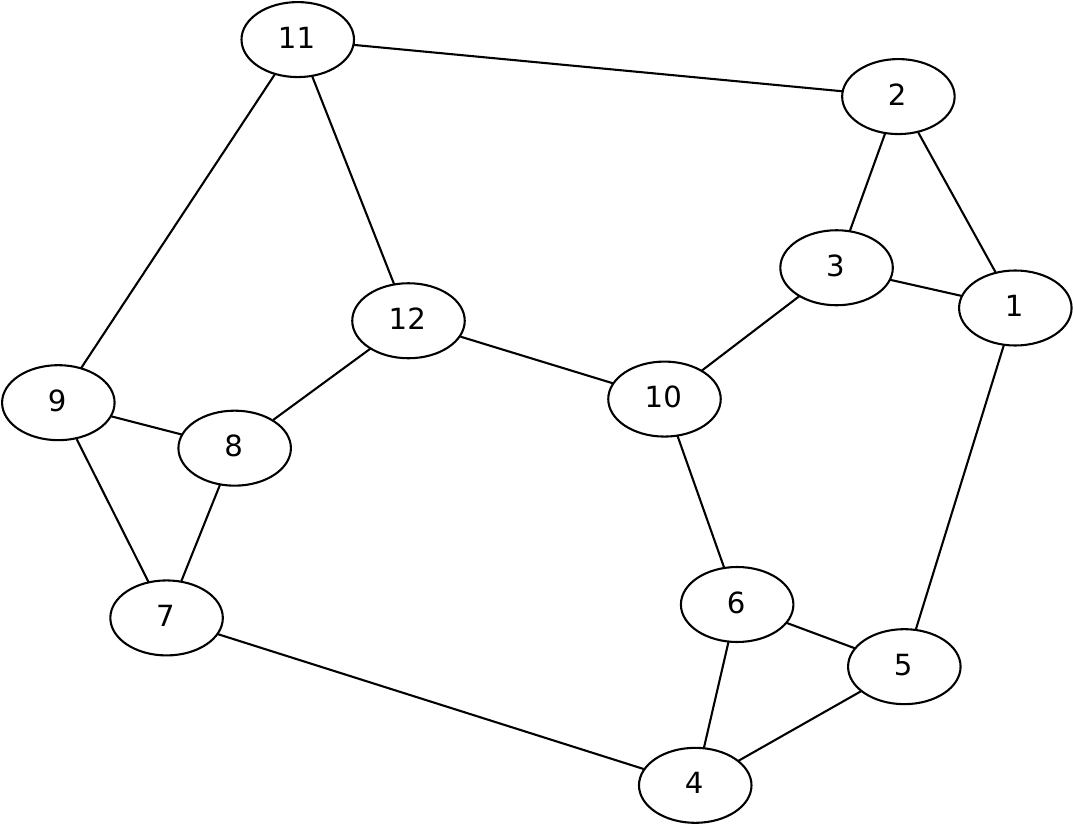}
    \end{minipage}
    \begin{minipage}[t]{0.45\textwidth}
        \centering
        \subfiguretitle{b)}
        \includegraphics[width=0.85\textwidth]{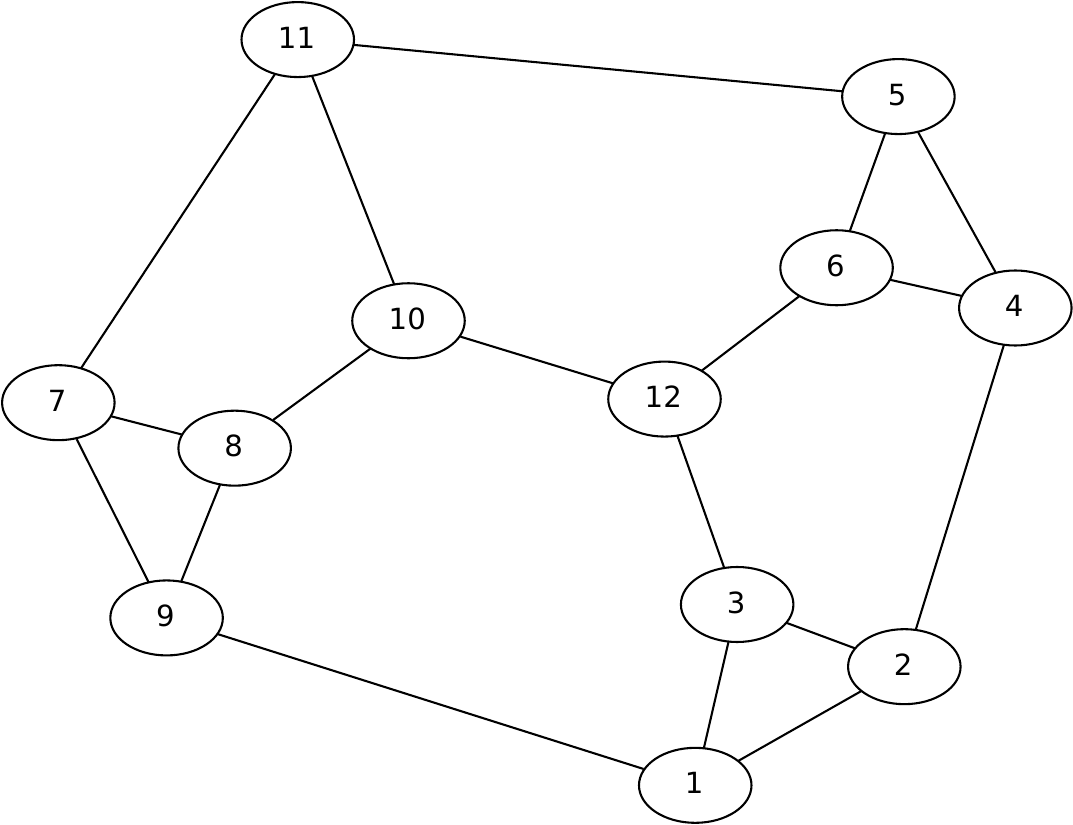}
    \end{minipage} \\[1ex]
    
    \begin{minipage}[t]{0.45\textwidth}
        \centering
        \subfiguretitle{c)}
        \includegraphics[width=0.75\textwidth]{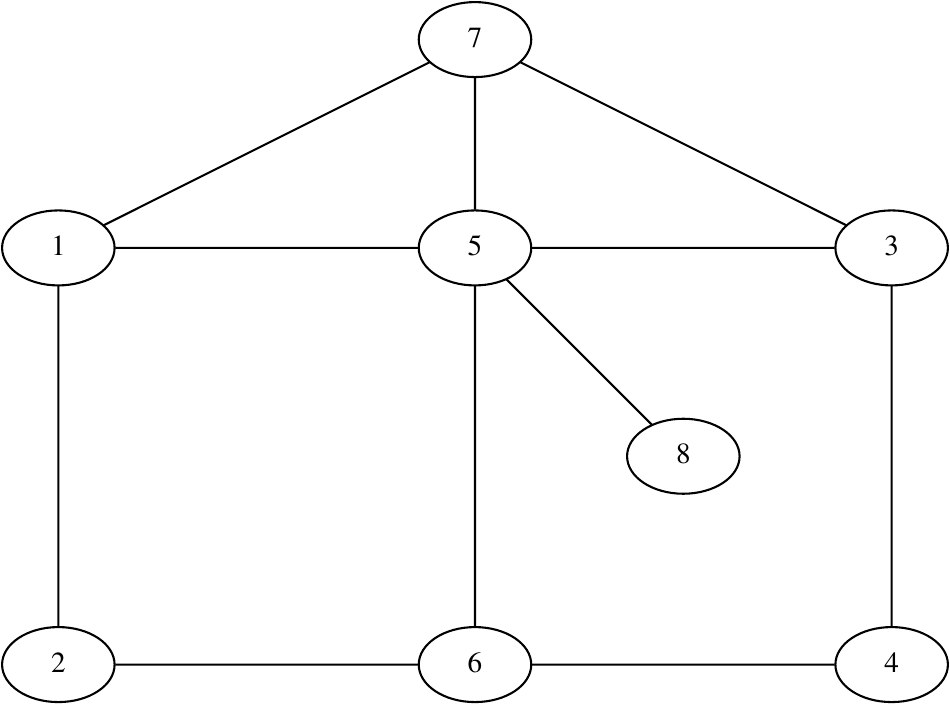}
    \end{minipage}
    \begin{minipage}[t]{0.45\textwidth}
        \centering
        \subfiguretitle{d)}
        \includegraphics[width=0.75\textwidth]{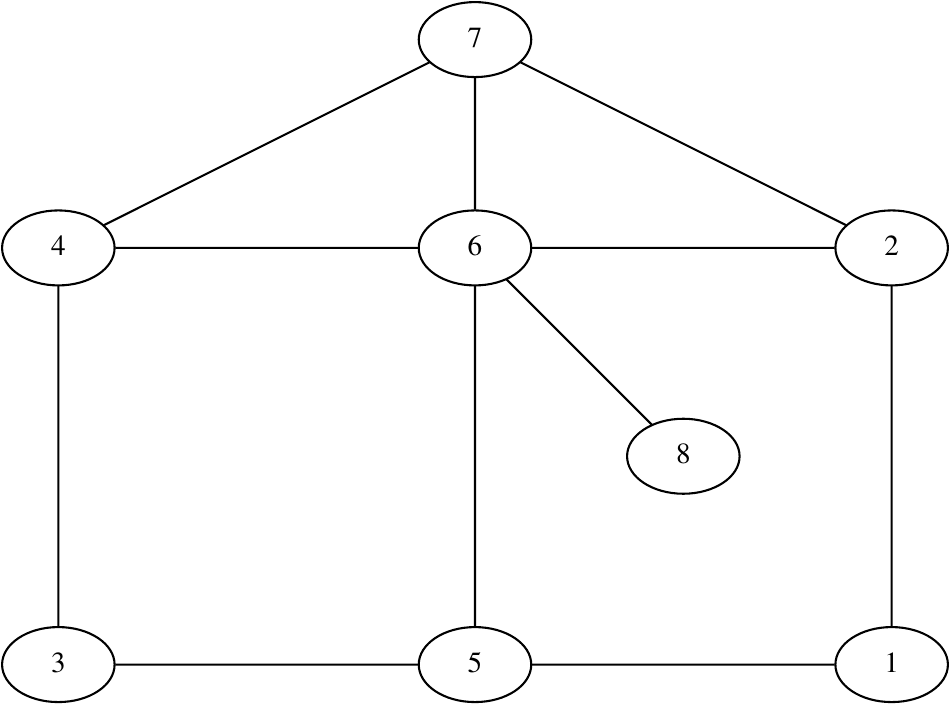}
    \end{minipage} \\[1ex]
    
    \begin{minipage}[t]{0.45\textwidth}
        \centering
        \subfiguretitle{e)}
        \includegraphics[width=0.725\textwidth]{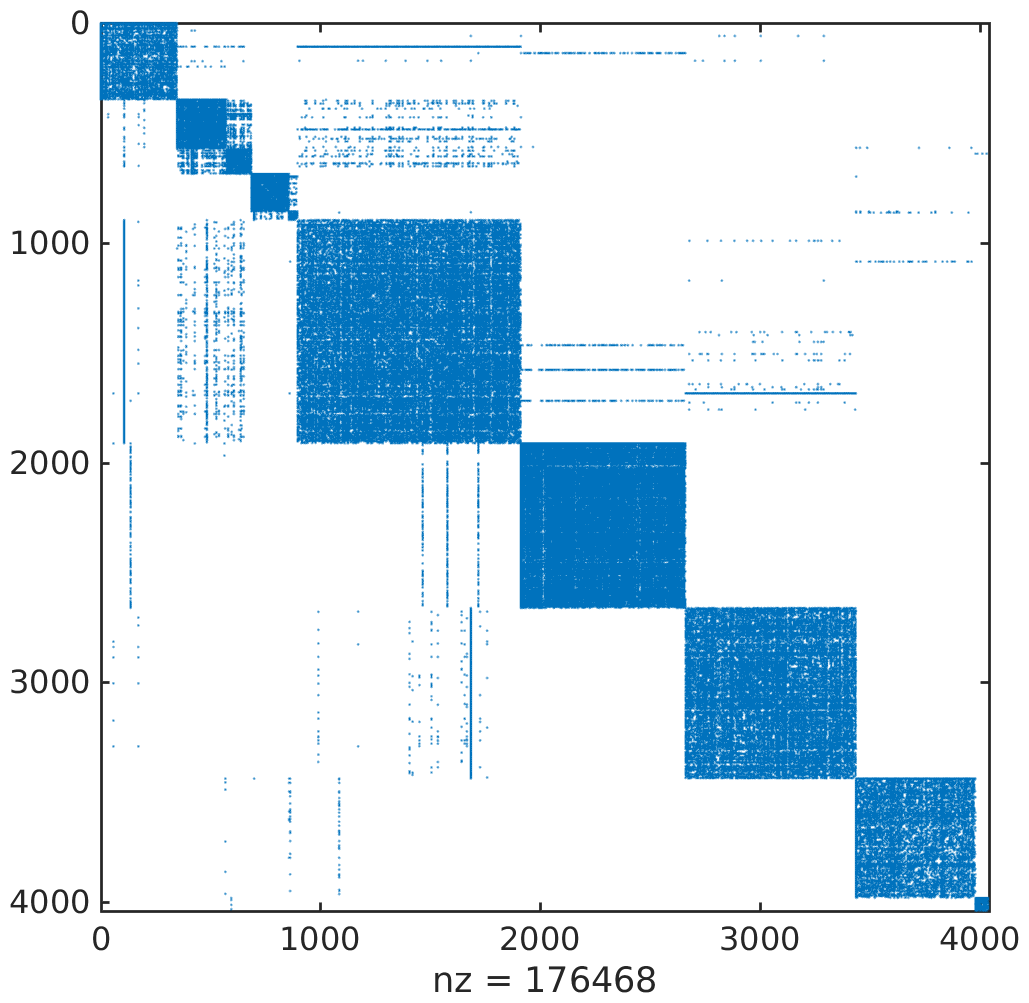}
    \end{minipage}
    \begin{minipage}[t]{0.45\textwidth}
        \centering
        \subfiguretitle{f)}
        \includegraphics[width=0.725\textwidth]{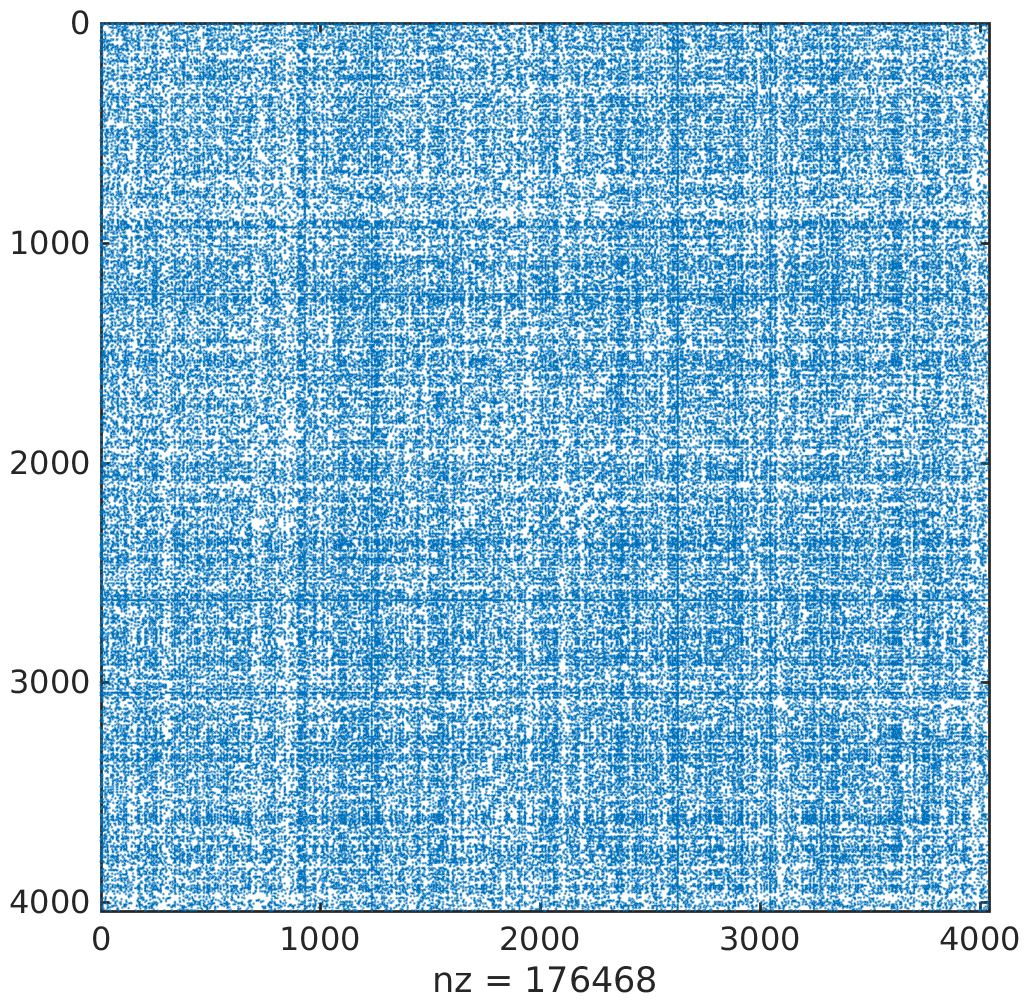}
    \end{minipage} \\[1ex]
    
    \begin{minipage}[t]{0.45\textwidth}
        \centering
        \subfiguretitle{g)}
        \includegraphics[width=0.7\textwidth]{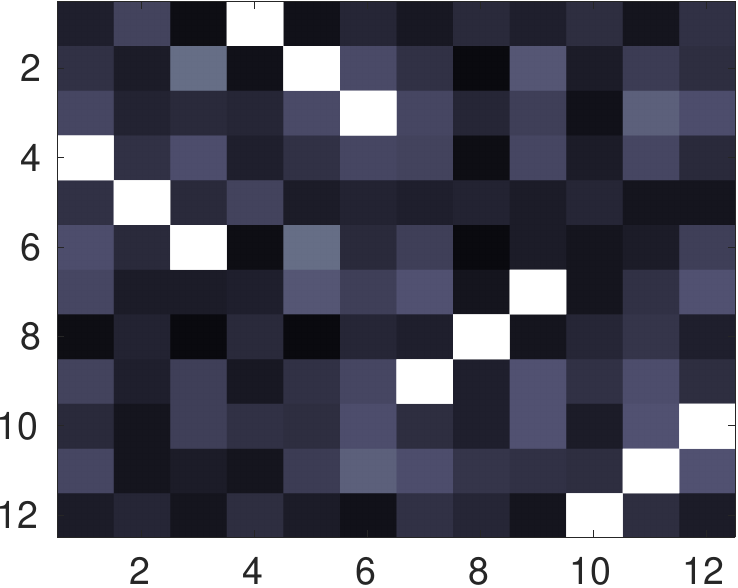}
    \end{minipage}
    \begin{minipage}[t]{0.45\textwidth}
        \centering
        \subfiguretitle{h)}
        \includegraphics[width=0.68\textwidth]{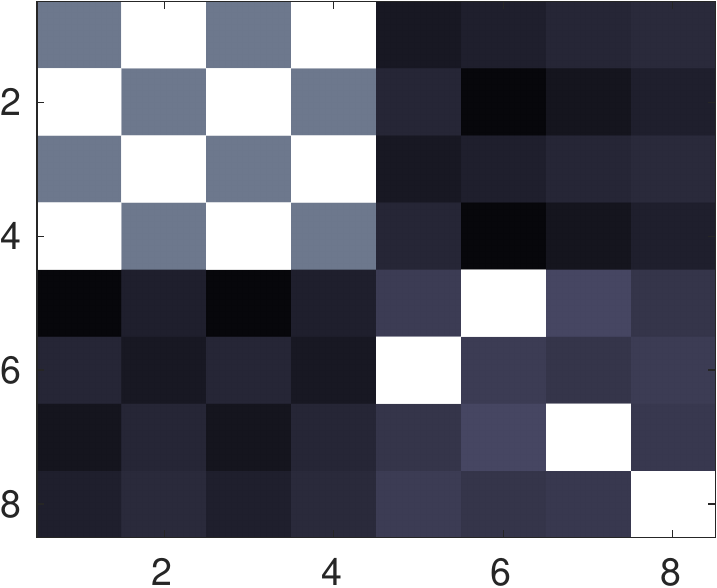}
    \end{minipage}
    \caption{a--b)~Original and permuted Frucht graph.
             c--d)~Original and permuted house graph.
             e--f)~Original and permuted adjacency matrix of the Facebook graph.
             g\protect\nobreakdash--h)~Cost matrix for Frucht graph and house graph. White entries represent possible assignments with cost $ c_{ij} < \varepsilon $.}
    \label{fig:Simple graphs}
\end{figure}

\end{example}

The examples show that even with incomplete information it is possible to compute valid solutions of the graph isomorphism problem. If the solution is not unique, constructing the cost matrix reduces the search space significantly since only zero-cost entries need to be taken into account. Furthermore, the examples illustrate the difficulties arising from ambiguous eigenvectors.

\section{Graphs with repeated eigenvalues}
\label{sec:Graphs with repeated eigenvalues}

If the graphs possess repeated eigenvalues, finding isomorphisms is much harder. The eigenvectors belonging to repeated eigenvalues are unique only up to basis rotations and we cannot construct a linear assignment problem by comparing corresponding eigenvectors anymore. As mentioned above, repeated eigenvalues typically correspond to graph symmetries. It can be shown that strongly regular graphs, for instance, possess at most three distinct eigenvalues~\cite{Lov07}.

\begin{definition} \label{def:Eigenvector partitioning}
Given a graph $ \mc{G}_A $ with adjacency matrix $ A $. Let $ \lambda_A = \big[ \lambda_A^{(1)}, \dots, \lambda_A^{(m)} \big] $ be the eigenvalues of the graph $ \mc{G}_A $ with multiplicities $ \mu_A = \big[ \mu_A^{(1)}, \dots, \mu_A^{(m)} \big] $, i.e., $ \sum_{k=1}^m \mu_A^{(k)} = n $. We then partition $ V_A $ into $ V_A = \big[ V_A^{(1)}, \dots, V_A^{(m)} \big] $, with $ V_A^{(k)} \in \R^{n \times \mu_A^{(k)}} $.
\end{definition}

That is, $ V_A^{(k)} $ is either the eigenvector belonging to the eigenvalue $ \lambda_A^{(k)} $ or the matrix whose columns form an orthogonal basis of the eigenspace.

\begin{example} \label{ex:Symmetric graphs}
We will use the following guiding examples to illustrate the proposed isomorphism testing approach for graphs with repeated eigenvalues:
\begin{enumerate}[leftmargin=0em,itemindent=1.7em,labelsep=0.3em,label=\roman*)] \setlength{\itemsep}{0mm}
\item The eigenvalues of the cycle graph shown in Figure~\ref{fig:Cycle graph}a are $ \lambda_A = [-2, -1, 1, 2] $ with multiplicities $ \mu_A = [1, 2, 2, 1] $. The automorphism group of the cycle graph is $ D_6 $ and thus $ \abs{\Aut(\mc{G})} = 12 $.

\item The eigenvalues of the strongly regular Paley graph shown in Figure~\ref{fig:Paley}a are $ \lambda_A = \big[ \frac{-1 - \sqrt{17}}{2}, \frac{-1 + \sqrt{17}}{2}, 8 \big] $ with multiplicities $ \mu_A = \big[ 8, 8, 1 \big] $. Here, $ \Aut(\mc{G}) \cong S_5 $. Thus, the graph possesses $ \abs{\Aut(\mc{G})} = 5! = 120 $ automorphisms. \exampleSymbol
\end{enumerate}
\end{example}

\subsection{Eigenpolytopes}

For the graphs in the previous example, it is not possible to apply Algorithm~\ref{alg:GI friendly graphs} directly. If we compare only eigenvectors belonging to distinct eigenvalues, this leads to $ C = 0 $. That is, any permutation matrix would be a feasible solution of the linear assignment problem. Therefore, we need to exploit additional information encoded in matrices representing orthogonal projections onto the eigenspace of repeated eigenvalues.

\begin{definition}
Let $ \lambda_A^{(k)} $ be a repeated eigenvalue of graph $ \mc{G}_A $. For a vertex $ \mc{v}_i $, define $ V_A^{(k)}(\mc{v}_i) $ to be the $ i $-th row of $ V_A^{(k)} $. The convex hull of all vectors $ V_A^{(k)}(\mc{v}_i) $, $ i = 1, \dots, n $, is called the \emph{eigenpolytope} of the graph belonging to the eigenvalue $ \lambda_A^{(k)} $.
\end{definition}

The row vectors $ V_A^{(k)}(\mc{v}_i) $ clearly depend on the orthogonal basis chosen for the eigenspace, but the scalar product is independent of the choice of basis~\cite{God98}. The matrix
\begin{equation*}
    E_A^{(k)} = V_A^{(k)} \big(V_A^{(k)}\big)^T,
\end{equation*}
i.e., $ (E_A^{(k)})_{ij} = \big\langle V_A^{(k)}(\mc{v}_i), \, V_A^{(k)}(\mc{v}_j) \big\rangle $, represents the orthogonal projection onto the column space of $ V_A^{(k)} $ and is an invariant of the eigenspace that does not depend on the orthogonal basis chosen for $ V_A^{(k)} $, see also~\cite{CG97}. Thus, $ E_B^{(k)} = P^T E_A^{(k)} P $, which in itself can again be interpreted as a graph isomorphism problem. For a detailed description of the relation between a graph and the geometry of its eigenpolytopes, we refer to~\cite{God98, CG97}. We now exploit properties of the matrices $ E_A^{(k)} $ to identify isomorphisms. In what follows, we will show that by comparing eigenvectors and eigenpolytopes, it is possible to compute isomorphisms of strongly regular graphs such as the Paley graph.

\subsection{A spectral assignment approach for graphs with symmetries}

As described above, repeated eigenvalues complicate graph isomorphism testing. Our heuristic approach is based on finding local perturbations of the adjacency matrices $ A $ and $ B $ that break symmetries without changing the isomorphism. Let us illustrate the basic idea with a simple example.

\begin{example} \label{ex:Cycle graph motivation}
Let us consider the cycle graphs shown in Figure~\ref{fig:Cycle graph}a. In order to find an assignment for vertices of $ \mc{G}_A $ to vertices of $ \mc{G}_B $, we perturb the adjacency matrices $ A $ and $ B $. If we add a self-loop to vertex $ \mc{v}_1 $ of $ \mc{G}_A $ and $ \mc{v}_1 $ of $ \mc{G}_B $, the two graphs remain isomorphic\footnote{Note that due to the cyclic symmetry, we could assign vertex $ \mc{v}_1 $ to any other vertex of $ \mc{G}_B $.}. Thus, we assign vertex $ \mc{v}_1 $ to vertex $ \mc{v}_1 $ of $ \mc{G}_B $. The updated graphs are shown in Figure~\ref{fig:Cycle graph}b, the red vertices denote self-loops with weight $ 1 $. Now, we try to assign vertex $ \mc{v}_2 $ of $ \mc{G}_A $ to a vertex of $ \mc{G}_B $. Since we have broken the cyclic symmetry, there are now only two possible assignments (due to the remaining reflection symmetry). Vertex $ \mc{v}_2 $ of $ \mc{G}_A $ could be either assigned to vertex $ \mc{v}_5 $ or $ \mc{v}_6 $ of $ \mc{G}_B $. Adding self-loops with weight $ 2 $ to vertex $ \mc{v}_2 $ of $ \mc{G}_A $ and vertex $ \mc{v}_5 $ of $ \mc{G}_B $ -- shown in Figure~\ref{fig:Cycle graph}c --, the resulting graph is friendly and thus asymmetric. The permutation matrix $ P $ could be computed using Algorithm~\ref{alg:GI friendly graphs}. Alternatively, the procedure described above can be repeated until a valid assignment for all vertices is found. The resulting graphs are shown in Figure~\ref{fig:Cycle graph}d. \exampleSymbol
\end{example}

Let us formalize the above procedure. We start with the original adjacency matrices $ A $ and $ B $ and construct cost matrices $ C^{(k)} $, $ k = 1, \dots, m $,  as follows. For simple eigenvectors, we use the cost matrix from Definition~\ref{def:Cost matrix}, i.e., $ C^{(k)} = C(V_A^{(k)}, V_B^{(k)}) $. For repeated eigenvalues, we compute the projection matrices $ E_A^{(k)} $ and $ E_B^{(k)} $ and check for each row $ i $ of $ E_A^{(k)} $ whether it can be written as a permutation of row $ j $ of $ E_B^{(k)} $ by comparing the sorted vectors\footnote{Assume that $ V_A^{(k)} $ and $ V_B^{(k)} $ are simple eigenvectors and contain the same entries, then comparing $ E_A^{(k)} $ and $ E_B^{(k)} $ leads to the same nonzero pattern as comparing the eigenvectors entry-wise.}.

\begin{definition}
Let $ E_A^{(k)}(\mc{v}_i) $ and $ E_B^{(k)}(\mc{v}_j) $ be the $ i $-th and $ j $-th row of the matrices $ E_A^{(k)} $ and $ E_B^{(k)} $, respectively, and let $ s : \R^n \to \R^n $ be a function that sorts the entries of a vector. For repeated eigenvalues, we define $ C^{(k)} = (c_{ij}^{(k)}) $, with $ c_{ij}^{(k)} = \big\lVert s \big( E_A^{(k)}(\mc{v}_i) \big) - s \big( E_B^{(k)}(\mc{v}_j) \big) \big\rVert_F $.
\end{definition}

Note that this is only a heuristic approach and might lead to wrong assignments. However, utilizing properties of the eigenpolytopes improves the efficiency of the algorithm significantly and backtracking is required only in exceptional cases. For two graphs $ \mc{G}_A $ and $ \mc{G}_B $, the cost matrix is then defined as
\begin{equation*}
    C(\mc{G}_A, \mc{G}_B) = \sum_{k=1}^m C^{(k)}.
\end{equation*}
The entries $ c_{ij} $ represent the costs of assigning $ \mc{v}_i $ of $ \mc{G}_A $ to $ \mc{v}_j $ of $ \mc{G}_B $. To determine possible assignments, we again compute $ C = C(\mc{G}_A, \mc{G}_B) $ and solve the resulting linear assignment problem
\begin{equation*}
    c = \min_{P \in \mathcal{P}_n} \tr \left (P^T C \right).
\end{equation*}

For the unperturbed cycle graph and Paley graph, the resulting cost matrices are zero, which implies that any vertex of $ \mc{G}_A $ can initially be assigned to any vertex of $ \mc{G}_B $. However, these assignment cannot be chosen independently. Thus, we assign vertices iteratively using local perturbations of the graphs as described in Example~\ref{ex:Cycle graph motivation}. After perturbing the graphs, symmetries are destroyed and the number of nonzero entries decreases until only one feasible solution remains. In order to perturb the adjacency matrices $ A $ and $ B $ and hence the eigenvalues and eigenvectors, we use single-entry matrices representing self-loops with different weights $ w $.

\begin{definition}
Define $ D_i(w) = \diag(d_1, \dots, d_n) $ to be the diagonal matrix with
\begin{equation*}
    d_j =
    \begin{cases}
        w, & \text{if } j = i, \\
        0, & \text{otherwise}.
    \end{cases}
\end{equation*}
\end{definition}

The proposed method for graphs with repeated eigenvalues is described in Algorithm~\ref{alg:GI symmetric graphs}. The algorithm can be stopped if the solution of the LAP is unique. The number of iterations required to obtain a unique solution depends on the order in which the vertices are perturbed. In the description of the algorithm, we have not included backtracking techniques. Backtracking is needed if a previously found assignment does not result in a correct permutation. We then delete the previous assignment and try to find a different assignment for the current vertex. Backtracking is required only for certain graph types as illustrated in Section~\ref{sec:Benchmark problems}.

\begin{algorithm}[htb]
    \caption{Graph isomorphism testing for graphs with repeated eigenvalues.}
    \label{alg:GI symmetric graphs}
    \vspace*{-0.5\baselineskip}
    \begin{enumerate} \setlength{\itemsep}{0mm}
        \item Compute $ C = C(\mc{G}_A, \mc{G}_B) $ and $ c = \min\limits_{P \in \mathcal{P}_n} \tr \left( P^T C \right) $.
        \item If $ \Lambda_A = \Lambda_B $ and $ c = 0 $, set $ i = 1 $.
        \item Define $ \widetilde{A} = A + D_i(i) $, $ \widetilde{B} = B + D_j(i) $, and $ \widetilde{C} = C(\mc{G}_{\widetilde{A}}, \mc{G}_{\widetilde{B}}) $.
        \item Find $ j $ so that $ \tilde{c} = \min\limits_{P \in \mathcal{P}_n} \tr \left( P^T \widetilde{C} \right) = 0 $.
        \item Set $ A = \widetilde{A} $, $ B = \widetilde{B} $. If $ i < n $, set $ i = i+1 $ and go to step 3.
    \end{enumerate}
    \vspace*{-0.5\baselineskip}
\end{algorithm}

\begin{example}
Let us consider again the graphs from Example~\ref{ex:Symmetric graphs}:
\begin{enumerate}[leftmargin=0em,itemindent=1.7em,labelsep=0.3em,label=\roman*)] \setlength{\itemsep}{0mm}
\item For the cycle graph, the cost matrices $ C $ that result in successful assignments are shown in Figure~\ref{fig:Cycle graph}a--d. Without perturbing the adjacency matrices, each vertex of $ \mc{G}_A $ can be assigned to each vertex of $ \mc{G}_B $ and the cost matrix $ C $ is zero. After one perturbation, all eigenvalues are distinct, but due to the remaining reflection symmetry the solution is not unique and there are still two ambiguous eigenvectors (see Example~\ref{ex:GI distint eigenvalues}). After two iterations, the solution is unique and the resulting permutation is given by
\begin{equation*}
    \pi =
    \begin{bmatrix}
        1 & 2 & 3 & 4 & 5 & 6 \\
        1 & 5 & 3 & 4 & 2 & 6
    \end{bmatrix}.
\end{equation*}

\begin{figure}[htb]
    \centering
    \begin{minipage}[c]{0.23\textwidth}
        \centering
        \subfiguretitle{a)}
        \includegraphics[width=0.9\textwidth]{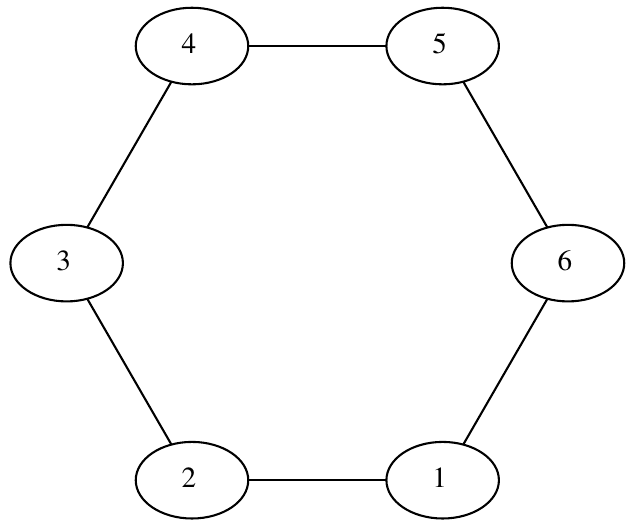}
    \end{minipage}
    \begin{minipage}[c]{0.23\textwidth}
        \centering
        \subfiguretitle{b)}
        \includegraphics[width=0.9\textwidth]{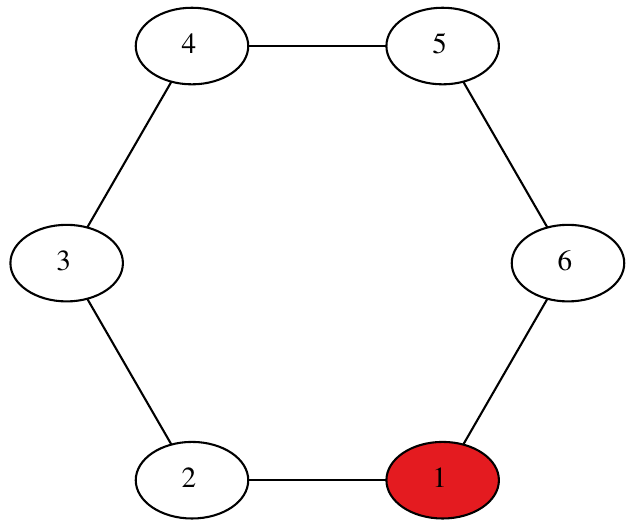}
    \end{minipage}
    \begin{minipage}[c]{0.23\textwidth}
        \centering
        \subfiguretitle{c)}
        \includegraphics[width=0.9\textwidth]{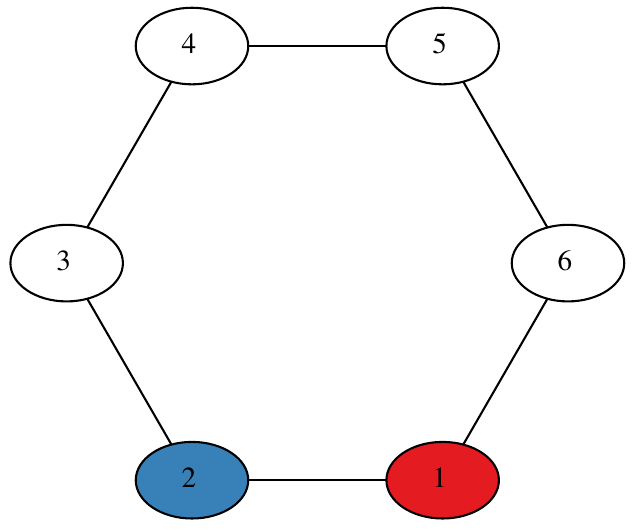}
    \end{minipage}
    \begin{minipage}[c]{0.03\textwidth}
        \vspace*{3ex} ...
    \end{minipage}
    \begin{minipage}[c]{0.23\textwidth}
        \centering
        \subfiguretitle{d)}
        \includegraphics[width=0.9\textwidth]{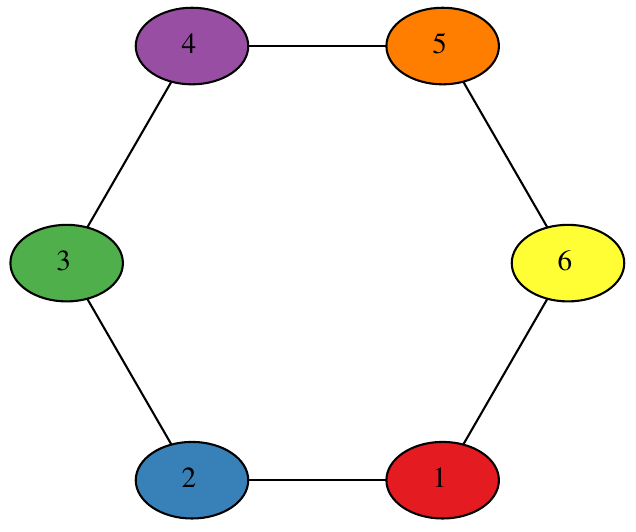}
    \end{minipage} \\[1ex]
    
    \begin{minipage}[c]{0.23\textwidth}
        \centering
        \includegraphics[width=0.9\textwidth]{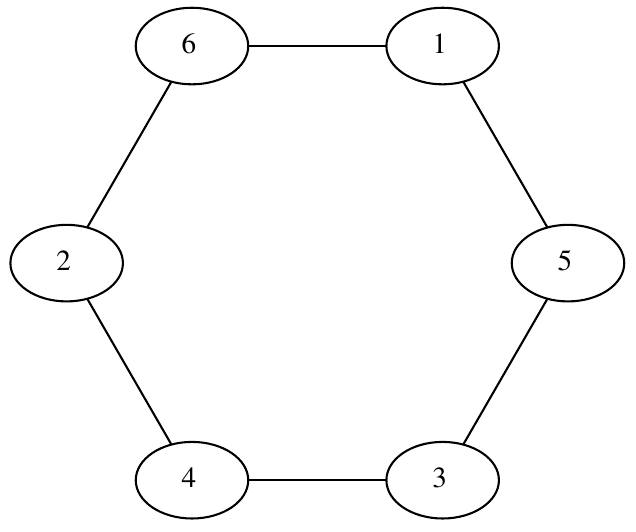}
    \end{minipage}
    \begin{minipage}[c]{0.23\textwidth}
        \centering
        \includegraphics[width=0.9\textwidth]{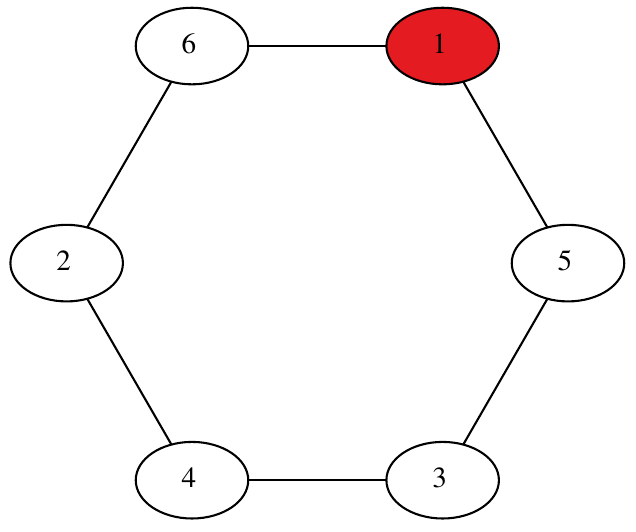}
    \end{minipage}
    \begin{minipage}[c]{0.23\textwidth}
        \centering
        \includegraphics[width=0.9\textwidth]{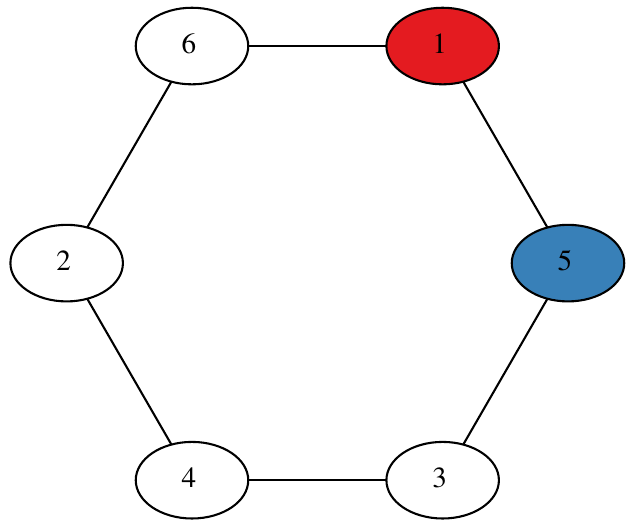}
    \end{minipage}
    \begin{minipage}[c]{0.03\textwidth}
        ...
    \end{minipage}
    \begin{minipage}[c]{0.23\textwidth}
        \centering
        \includegraphics[width=0.9\textwidth]{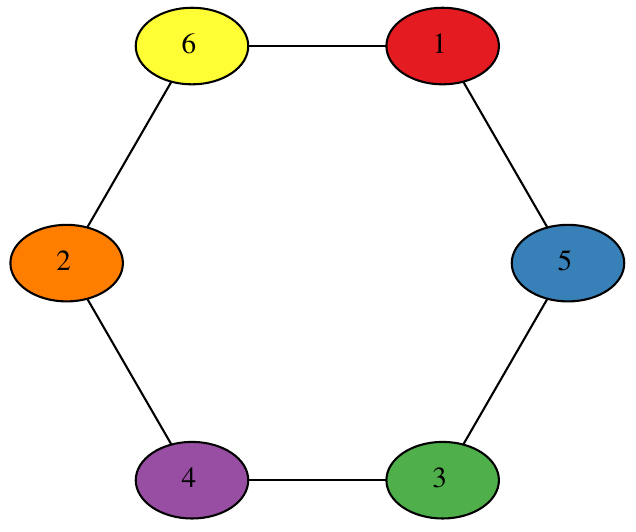}
    \end{minipage} \\[1.5ex]
    
    \begin{minipage}[c]{0.23\textwidth}
        \centering
        \includegraphics[width=0.8\textwidth]{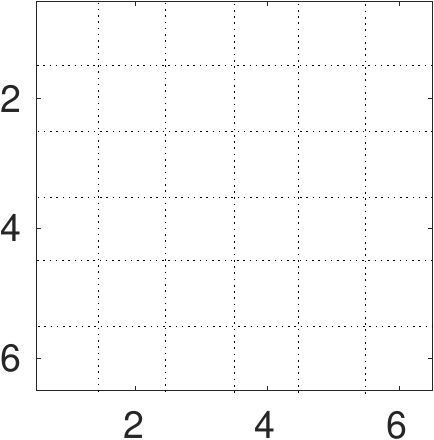}
    \end{minipage}
    \begin{minipage}[c]{0.23\textwidth}
        \centering
        \includegraphics[width=0.8\textwidth]{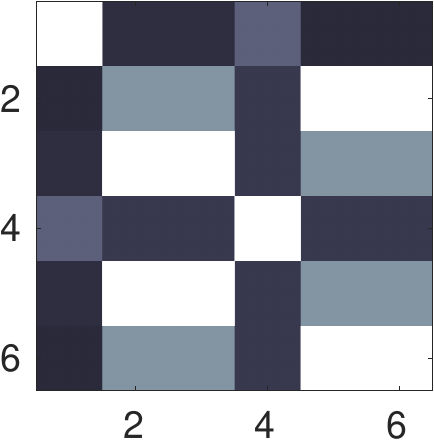}
    \end{minipage}
    \begin{minipage}[c]{0.23\textwidth}
        \centering
        \includegraphics[width=0.8\textwidth]{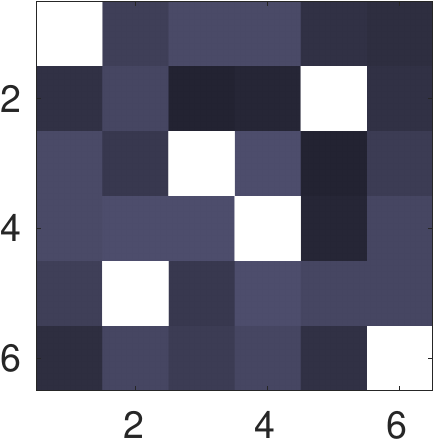}
    \end{minipage}
    \begin{minipage}[c]{0.03\textwidth}
        ...
    \end{minipage}
    \begin{minipage}[c]{0.23\textwidth}
        \centering
        \includegraphics[width=0.8\textwidth]{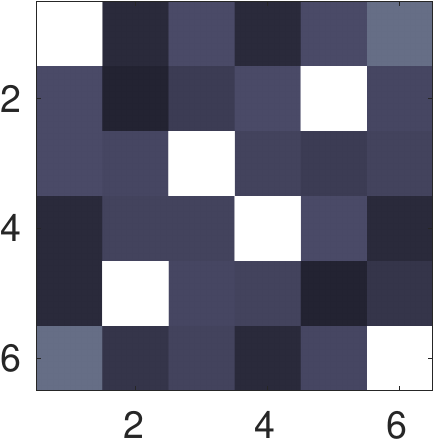}
    \end{minipage}   
    \caption{Graph isomorphism testing procedure for the cycle graph. The various colors represent self-loops with different weights. The bottom row shows the structure of the corresponding cost matrices $ C $. White entries represent possible assignments with cost $ c_{ij} < \varepsilon $. After two perturbations of the graphs, the solution of the LAP is unique.}
    \label{fig:Cycle graph}
\end{figure}

\item For the Paley graph, the cost matrices after 1, 2, 3, and 4 successful assignments are shown in Figure~\ref{fig:Paley}b--e. After the fourth perturbation, the solution is unique and
\begin{equation*}
    \pi =
    \begin{bmatrix}
         1 &  2 &  3 &  4 &  5 &  6 &  7 &  8 &  9 & 10 & 11 & 12 & 13 & 14 & 15 & 16 & 17 \\
         1 &  6 & 15 &  3 & 11 &  7 & 17 & 12 &  9 &  8 &  4 &  2 &  5 & 16 & 10 & 13 & 14
    \end{bmatrix}. \tag*{\exampleSymbol}
\end{equation*}

\begin{figure}[htb]
    \centering
    \begin{minipage}[r]{0.24\textwidth}
        \centering
        \subfiguretitle{b)}
        \includegraphics[width=\textwidth]{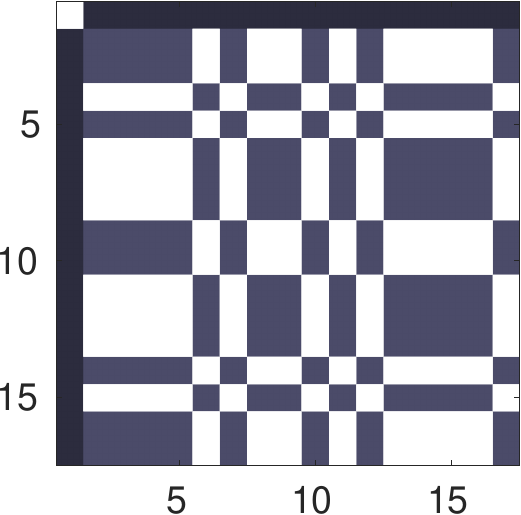}
        \subfiguretitle{c)}
        \includegraphics[width=\textwidth]{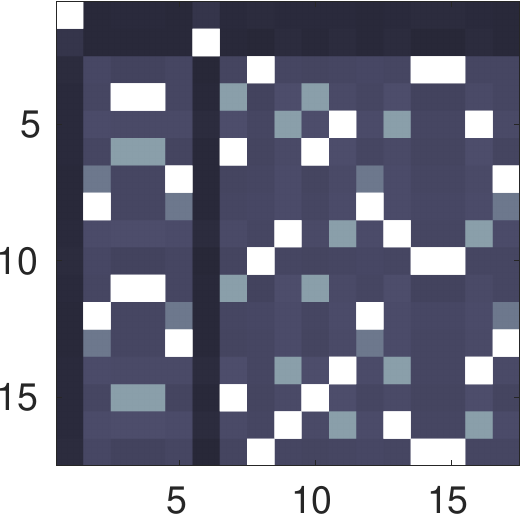}
    \end{minipage}
    \begin{minipage}[c]{0.5\textwidth}
        \centering
        \subfiguretitle{a)}
        \includegraphics[width=\textwidth]{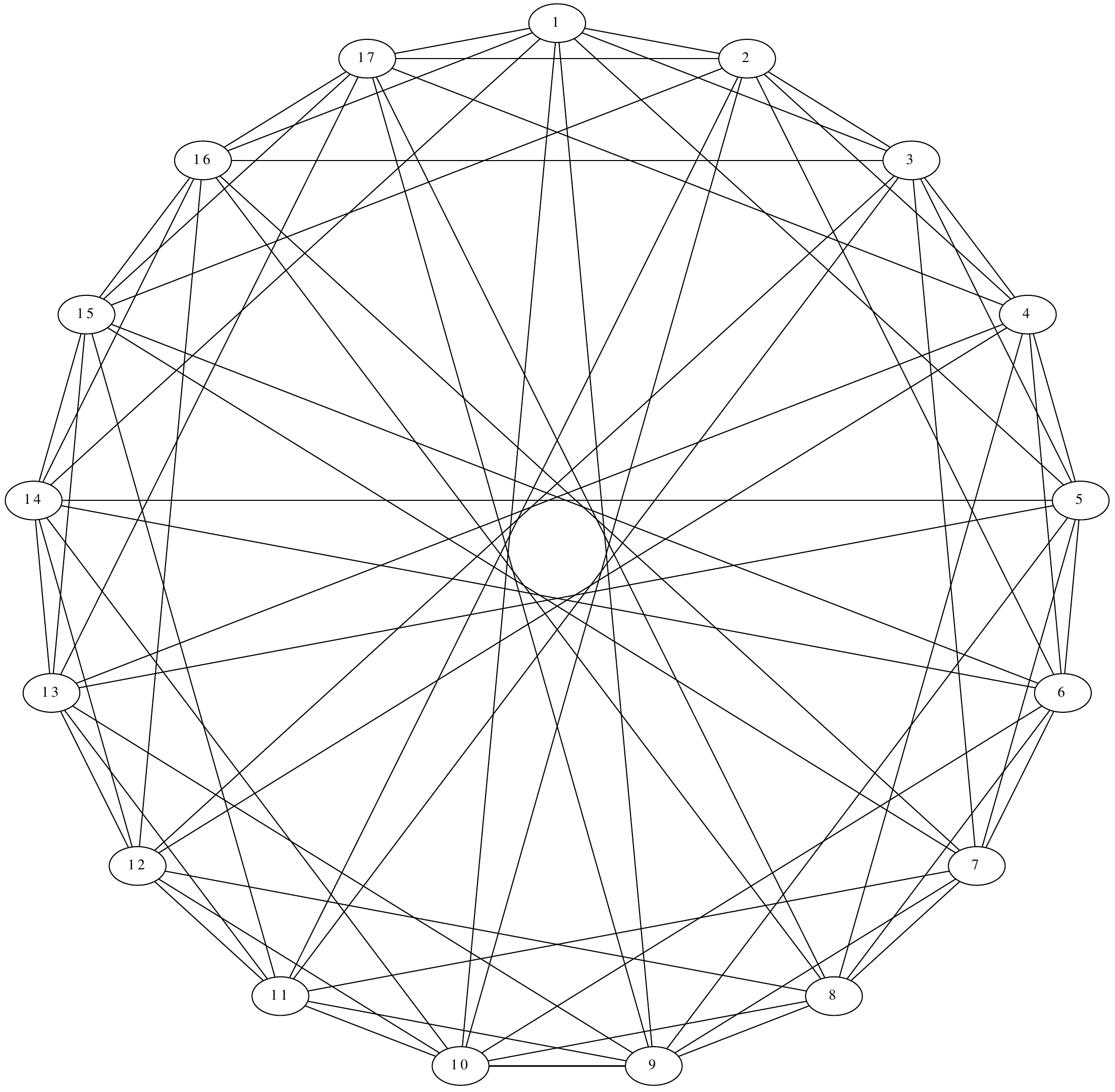}
    \end{minipage}
    \begin{minipage}[c]{0.24\textwidth}
        \centering
        \subfiguretitle{d)}
        \includegraphics[width=\textwidth]{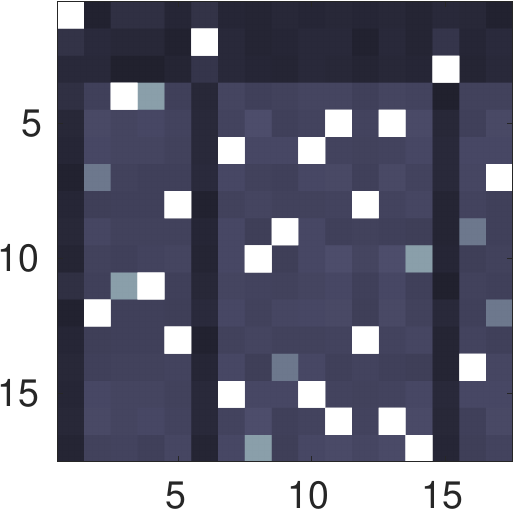}
        \subfiguretitle{e)}
        \includegraphics[width=\textwidth]{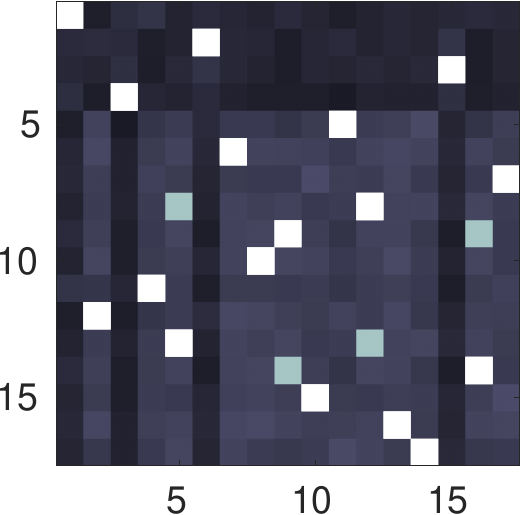}
    \end{minipage}
    \caption{a) Strongly regular Paley graph. b--e) Structure of the cost matrices $ C $ after 1, 2, 3, and 4 successful assignments. White entries represent assignments with $ c_{ij} < \varepsilon $. After four perturbations, the solution is unique.}
    \label{fig:Paley}
\end{figure}
\end{enumerate}

\end{example}

These examples demonstrate that the proposed method successfully computes isomorphisms of graphs with repeated eigenvalues.

\section{Benchmark problems}
\label{sec:Benchmark problems}

In this section, we will present numerical results for benchmark problems downloaded from~\cite{Val11} and~\cite{LAC11a}. For our computations, we used \emph{Matlab} and an error tolerance $ \varepsilon = 10^{-6} $. That is, two eigenvalues of a matrix are defined to be identical if the difference is less than $ \varepsilon $. Furthermore, an assignment is accepted if the cost $ c $ of the solution of the LAP is less than $ \varepsilon $. For all graphs downloaded from~\cite{Val11}, the algorithm returned correct results without backtracking. Results for larger benchmark problems used by \emph{conauto} are presented in Table~\ref{tab:BenchmarkResults}. For each benchmark problem, we run the proposed algorithm 100 times using different randomly generated permutations of the original graph. Here, \emph{n} is the number of vertices, \emph{noBT} the number of runs where no backtracking was required, and \emph{BT} the number of runs where backtracking was required to find an isomorphism. The column \emph{steps} describes the average number of backtracking steps needed to find an isomorphism and the last column lists the average runtime in seconds. The efficiency of the algorithm could be easily improved by using C or C\texttt{++}. The results show that the algorithm returns correct results for most of the benchmark problems without backtracking. Only the Steiner triple system graph (1) and the union of strongly regular graphs (16) require backtracking for almost all test cases.

\begin{table}[htb]
    \newcommand{\muco}[1]{\multicolumn{1}{|c|}{#1}}
    \centering
    \caption{Test results for various benchmark graphs.}
    \scriptsize
    \begin{tabular}{|l|r|l|r|r|r|r|r|}
        \hline
        \muco{Type} & \muco{\#} & \muco{Name} & $ n $ & \muco{noBT} & \muco{BT} & \muco{steps (avg)} & \muco{time [s]} \\
        \hline
        \hline
        \multirow{1}{*}{Steiner triple system graphs}                           &  1 & sts-19\_57         & 57 &   5 &  95 & 37.91 & 11.051 \\ \hline 
        \multirow{3}{*}{\pbox{5cm}{Latin square graphs \\ (prime order)}}       &  2 & latin-3\_9         &  9 & 100 &   0 &  0.00 &  0.004 \\ 
                                                                                &  3 & latin-5\_25        & 25 & 100 &   0 &  0.00 &  0.017 \\
                                                                                &  4 & latin-7\_49        & 49 &  98 &   2 &  1.00 &  0.141 \\ \hline
        \multirow{3}{*}{\pbox{5cm}{Latin square graphs \\ (prime power order)}} &  5 & latin-2\_4         &  4 & 100 &   0 &  0.00 &  0.001 \\ 
                                                                                &  6 & latin-4\_16        & 16 & 100 &   0 &  0.00 &  0.008 \\
                                                                                &  7 & latin-6\_36        & 36 &  74 &  26 &  2.31 &  0.086 \\ \hline
        Paley graphs                                                            &  8 & paley-prime\_13    & 13 &  89 &  11 &  1.09 &  0.006 \\ 
        (prime order)                                                           &  9 & paley-prime\_29    & 29 & 100 &   0 &  0.00 &  0.031 \\ \hline
        Paley graphs                                                            & 10 & paley-power\_9     &  9 & 100 &   0 &  0.00 &  0.002 \\ 
        (prime power order)                                                     & 11 & paley-power\_25    & 25 & 100 &   0 &  0.00 &  0.020 \\ \hline                     
        \multirow{2}{*}{Lattice graphs}                                         & 12 & lattice(4)\_16     & 16 & 100 &   0 &  0.00 &  0.010 \\ 
                                                                                & 13 & lattice(6)\_36     & 36 & 100 &   0 &  0.00 &  0.067 \\ \hline
        \multirow{2}{*}{Triangular graphs}                                      & 14 & triangular(7)\_21  & 21 & 100 &   0 &  0.00 &  0.013 \\ 
                                                                                & 15 & triangular(10)\_45 & 45 & 100 &   0 &  0.00 &  0.097 \\ \hline
        Unions of strongly regular graphs                                       & 16 & usr(1)\_29-1       & 29 &  11 &  89 & 13.49 &  0.392 \\ \hline 
        Clique-conected cubic                                                   & 17 & chh\_cc(1-1)\_22-1 & 22 & 100 &   0 &  0.00 &  0.006 \\ 
        hypo-Hamiltonian graphs                                                 & 18 & chh\_cc(2-1)\_44-1 & 44 & 100 &   0 &  0.00 &  0.097 \\ \hline
        Non-disjoint unions of                                                  & 19 & tnn(1)\_26-1       & 26 & 100 &   0 &  0.00 &  0.066 \\ 
        undirected tripartite graphs                                            & 20 & tnn(2)\_52-1       & 52 & 100 &   0 &  0.00 &  0.700 \\ \hline
        \multirow{2}{*}{Random graphs}                                          & 21 & iso\_r01N\_s20     & 20 & 100 &   0 &  0.00 &  0.002 \\ 
                                                                                & 22 & iso\_r01N\_s40     & 40 & 100 &   0 &  0.00 &  0.010 \\
        \hline
    \end{tabular}
    \label{tab:BenchmarkResults}
\end{table}

\begin{figure}[htb]
    \centering
    \includegraphics[width=0.95\textwidth]{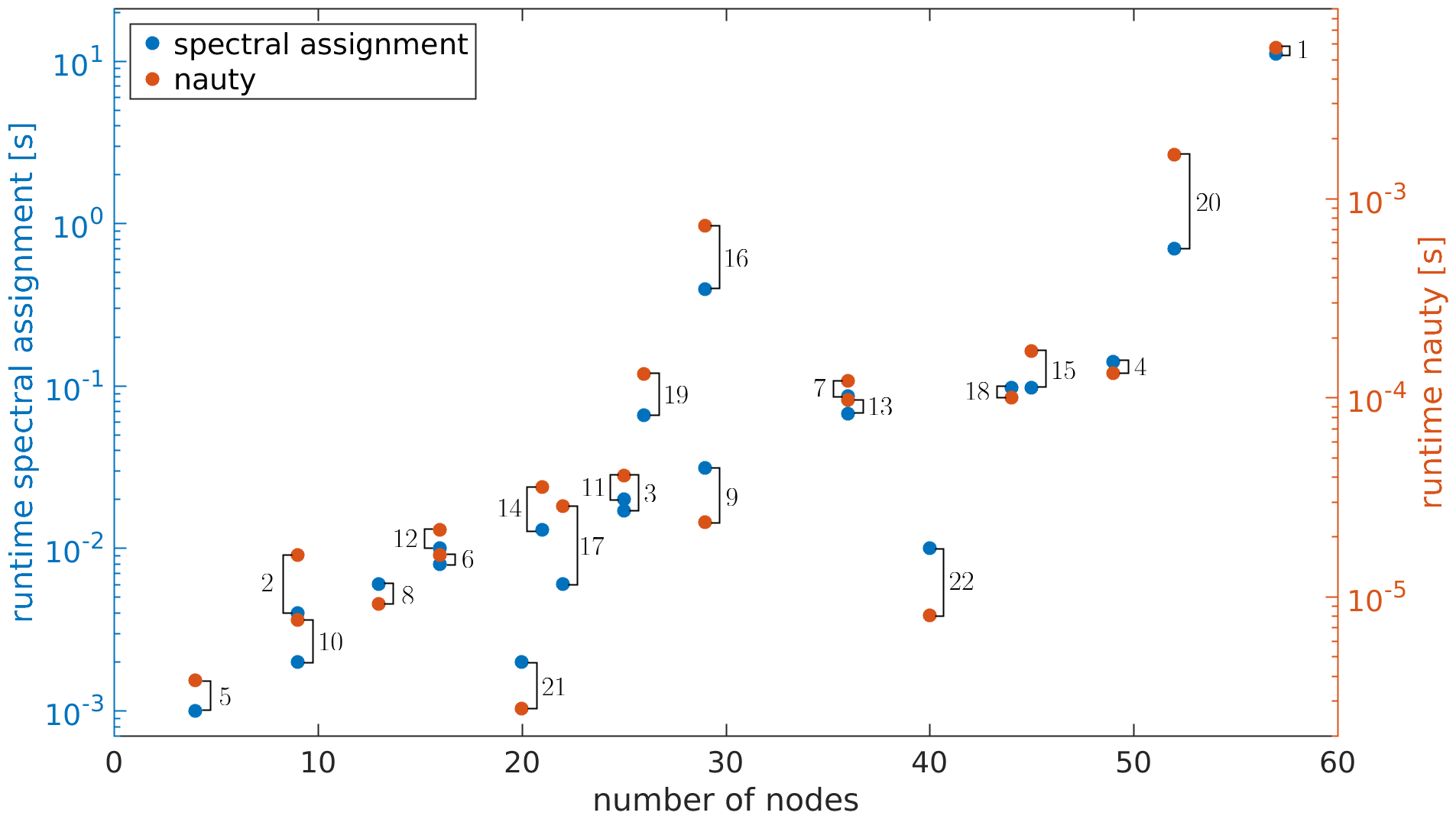}
    \caption{Comparison of the runtimes of the spectral assignment approach and \emph{nauty}. Note that two different axes are used.}
    \label{fig:Runtimes}
\end{figure}

In order to analyze the scalability of the spectral assignment approach, we compare it with the state-of-the-art graph automorphism and isomorphism tool \emph{nauty}~\cite{MK81, MKP14}. For each benchmark graph, we run \emph{nauty} 100 times using different randomly generated permutations. Additionally, each GI instance is solved 10000 times to obtain more accurate runtimes. The results are shown in Figure~\ref{fig:Runtimes}. We expect similar results for other tools such as \emph{conauto}~\cite{LAC11b} or \emph{bliss}~\cite{JK07} (for a comparison of these algorithms, see~\cite{MKP14}). While the absolute runtimes of \emph{nauty}, which is implemented in C, are much lower than the runtimes of our proof-of-concept Matlab implementation, the complexity of the spectral assignment grows only slightly faster. Furthermore, the comparison shows that in particular the random graphs (21) and (22) seem to be comparably easy to solve, while the union of strongly regular graphs (16) and the Steiner triple system graph (1) seem particularly hard to solve for both nauty and spectral approaches. This is also reflected in the number of backtracking steps. The spectral assignment approach for graphs with repeated eigenvalues could be optimized by a more sophisticated assignment strategy and by combining it with other heuristics. Instead of assigning nodes depending on the node numbers as described in Algorithm~\ref{alg:GI symmetric graphs}, it might be more efficient to exploit properties of the graph to decide which node should be assigned next. This is expected to reduce the number of backtracking steps and will be the focus of our future work.

\section{Conclusion}
\label{sec:Conclusion}

In this work, we have presented eigendecomposition-based methods for solving the graph isomorphism problem. The algorithms were demonstrated with the aid of several guiding examples and benchmark problems. For friendly graphs, we have proven that the problem can be cast as a linear assignment problem. The approach was then generalized to unambiguous graphs. The examples show that the assignment problem formulation results in correct solutions even for ambiguous graphs. The primary issue related to the influence of ambiguous eigenvectors is the number of automorphisms and feasible solutions of the LAP. For graphs with repeated eigenvalues, our approach relies on the repeated perturbation of the adjacency matrices and solution of linear assignment problems. By exploiting properties of eigenpolytopes, it is possible to check whether two highly symmetric graphs are isomorphic. We believe that the proposed approach can be used to efficiently find isomorphisms, to detect and break symmetries, and to gain insight into the structure of highly regular graphs. Other properties of the eigenpolytopes may be exploited to minimize the number of erroneous assignments which then require backtracking. An important open question is the classification of graphs that require backtracking in the spectral approach and ones that do not. The isomorphisms for graphs that do not require backtracking can consequently be computed in polynomial time. We conjecture that graphs that requires backtracking have additional structure that makes the computations particularly challenging.

In practical applications, the graphs might be contaminated by noise \cite{ABK15}. Instead of finding a perfect matching with zero assignment cost, the goal then is to find a permutation which minimizes a given cost function. This is also called the \emph{inexact} graph isomorphism problem. Future work includes investigating whether our approach can also be used for the inexact problem formulation. Since the eigenvalues of a graph depend continuously on the entries of the adjacency matrix, a slightly perturbed graph will have a similar spectrum. Thus, instead of determining whether two graphs are isomorphic, the assignment approach can potentially be generalized so that the best matching of two graphs is computed, i.e., a permutation that minimizes the Frobenius norm distance between them. We believe that the Frobenius norm will serve as a good cost function for the inexact isomorphism problem. If the noise, however, is large, the spectrum of the graph might change in such a way that it becomes impossible to compare corresponding eigenvalues and eigenvectors.

\section*{Acknowledgements}

We would like to thank the reviewers for their helpful comments and suggestions.

\bibliographystyle{unsrt}
\bibliography{GI}

\begin{thebibliography}{10}

\bibitem{ABK15}
Y.~Aflalo, A.~Bronstein, and R.~Kimmel.
\newblock On convex relaxation of graph isomorphism.
\newblock {\em Proceedings of the National Academy of Sciences},
  112(10):2942--2947, 2015.

\bibitem{Koe06}
J.~K{\"o}bler.
\newblock On graph isomorphism for restricted graph classes.
\newblock In A.~Beckmann, U.~Berger, B.~L{\"o}we, and J.~Tucker, editors, {\em
  Logical Approaches to Computational Barriers}, Lecture Notes in Computer
  Science, pages 241--256. Springer, 2006.

\bibitem{Bab15}
L{\'{a}}szl{\'{o}} Babai.
\newblock Graph isomorphism in quasipolynomial time.
\newblock {\em CoRR}, abs/1512.03547, 2015.

\bibitem{VT05}
V.~Arvind and J.~Tor{\'a}n.
\newblock Isomorphism testing: Perspective and open problems.
\newblock {\em Bulletin of the European Association for Theoretical Computer
  Science}, 86:66--84, 2005.

\bibitem{HRT03}
S.~Hallgren, A.~Russell, and A.~Ta-Shma.
\newblock The hidden subgroup problem and quantum computation using group
  representations.
\newblock {\em SIAM Journal on Computing}, 32(4):916--934, 2003.

\bibitem{HT72}
J.~E. Hopcroft and R.~E. Tarjan.
\newblock Isomorphism of planar graphs.
\newblock In R.~E. Miller and J.~W. Thatcher, editors, {\em Complexity of
  Computer Computations}, pages 131--152. Plenum Press, 1972.

\bibitem{Luk82}
E.~M. Luks.
\newblock Isomorphism of graphs of bounded valence can be tested in polynomial
  time.
\newblock {\em Journal of Computer and System Sciences}, 25(1):42--65, 1982.

\bibitem{BGM82}
L.~Babai, D.~Y. Grigoryev, and D.~M. Mount.
\newblock Isomorphism of graphs with bounded eigenvalue multiplicity.
\newblock In {\em Proceedings of the 14th Annual ACM Symposium on Theory of
  Computing}, pages 310--324. ACM, 1982.

\bibitem{FS15}
M.~Fiori and G.~Sapiro.
\newblock On spectral properties for graph matching and graph isomorphism
  problems.
\newblock {\em Information and Inference}, 4(1):63--76, 2015.

\bibitem{Spi09}
D.~A. Spielman.
\newblock Spectral graph theory ({L}ecture notes).
\newblock \url{http://www.cs.yale.edu/homes/spielman}, 2009.

\bibitem{Spi96}
D.~A. Spielman.
\newblock Faster isomorphism testing of strongly regular graphs.
\newblock In {\em Proceedings of the 28th Annual ACM Symposium on Theory of
  Computing}, pages 576--584. ACM, 1996.

\bibitem{Lov07}
L.~Lov{\'a}sz.
\newblock Eigenvalues of graphs.
\newblock Technical report, E{\"o}tv{\"o}s Lor{\'a}nd University, Budapest,
  Hungary, 2007.

\bibitem{LM79}
F.~T. Leighton and G.~L. Miller.
\newblock Certificates for graphs with distinct eigenvalues.
\newblock Orginal Manuscript, 1979.

\bibitem{Sch68}
P.~Sch{\"o}nemann.
\newblock On two-sided orthogonal {P}rocrustes problems.
\newblock {\em Psychometrika}, 33(1):19--33, 1968.

\bibitem{GD04}
J.~C. Gower and G.~B. Dijksterhuis.
\newblock {\em Procrustes Problems}.
\newblock Number~30 in Oxford statistical science series. Oxford University
  Press, 2004.

\bibitem{OB12}
I.~Oren and R.~Band.
\newblock Isospectral graphs with identical nodal counts.
\newblock {\em Journal of Physics A: Mathematical and Theoretical},
  45(13):1--11, 2012.

\bibitem{Kuh55}
H.~W. Kuhn.
\newblock The {H}ungarian method for the assignment problem.
\newblock {\em Naval Research Logistics Quarterly}, 2:83--97, 1955.

\bibitem{BC99}
R.~E. Burkard and E.~\c{C}ela.
\newblock Linear assignment problems and extensions.
\newblock In D.-Z. Du and P.~M. Pardalos, editors, {\em Handbook of
  Combinatorial Optimization}. Kluwer, 1999.

\bibitem{ML12}
J.~McAuley and J.~Leskovec.
\newblock Learning to discover social circles in ego networks.
\newblock {\em Advances in Neural Information Processing Systems 25}, pages
  539--547, 2012.

\bibitem{LK14}
J.~Leskovec and A.~Krevl.
\newblock {SNAP Datasets}: {Stanford} large network dataset collection.
\newblock \url{http://snap.stanford.edu/data}, 2014.

\bibitem{God98}
C.~D. Godsil.
\newblock Eigenpolytopes of distance regular graphs.
\newblock {\em Canadian Journal of Mathematics}, 50(4):739--755, 1998.

\bibitem{CG97}
A.~Chan and C.~D. Godsil.
\newblock Symmetry and eigenvectors.
\newblock In G.~Hahn and G.~Sabidussi, editors, {\em Graph Symmetry: Algebraic
  Methods and Applications}. Kluwer, 1997.

\bibitem{Val11}
V.~V. Valiayeu.
\newblock \emph{Griso} for regular graphs.
\newblock \url{http://sourceforge.net/projects/griso}, 2011.

\bibitem{LAC11a}
J.~L. L\'opez-Presa, A.~{Fern{\'a}ndez Anta}, and L.~{N\'u\~nez Chiroque}.
\newblock Graph isomorphism algorithm \emph{conauto}.
\newblock https://sites.google.com/site/giconauto, 2011.

\bibitem{MK81}
Brendan~D. McKay.
\newblock Practical graph isomorphism, 1981.

\bibitem{MKP14}
B.~D. McKay and A.~Piperno.
\newblock Practical graph isomorphism, ii.
\newblock {\em Journal of Symbolic Computation}, 60:94--112, 2014.

\bibitem{LAC11b}
J.~L. L\'opez-Presa, A.~{Fern{\'a}ndez Anta}, and L.~{N\'u\~nez Chiroque}.
\newblock Conauto-2.0: Fast isomorphism testing and automorphism group
  computation.
\newblock {\em ArXiv e-prints}, 2011.

\bibitem{JK07}
T.~Junttila and P.~Kaski.
\newblock Engineering an efficient canonical labeling tool for large and sparse
  graphs.
\newblock In D.~Applegate, G.~St{\o}lting Brodal, D.~Panario, and R.~Sedgewick,
  editors, {\em Proceedings of the Ninth Workshop on Algorithm Engineering and
  Experiments and the Fourth Workshop on Analytic Algorithms and
  Combinatorics}, pages 135--149. SIAM, 2007.

\end{thebibliography}

\end{document}